\documentclass[sn-mathphys,Numbered]{sn-jnl}

\usepackage{graphicx}%
\usepackage{multirow}%
\usepackage{amsmath,amssymb,amsfonts}%
\usepackage{amsthm}%
\usepackage{mathrsfs}%
\usepackage[title]{appendix}%
\usepackage{xcolor}%
\usepackage{textcomp}%
\usepackage{manyfoot}%
\usepackage{booktabs}%
\usepackage{algorithm}%
\usepackage{algorithmicx}%
\usepackage{algpseudocode}%
\usepackage{listings}
\usepackage{subcaption}

\theoremstyle{thmstyleone}%
\theoremstyle{thmstyletwo}%

\theoremstyle{thmstylethree}%

\newcommand{\new}[1]{{\color{black}#1}}

\begin{document}
	\title[Universal transition of spectral fluctuation in particle-hole symmetric system]{Universal transition of spectral fluctuation in particle-hole symmetric system}
	\author*[1]{\fnm{Triparna} \sur{Mondal}} \email{t.mondal@vecc.gov.in}
	\author[1,2]{\fnm{Shashi C. L.} \sur{Srivastava}} 
	\affil*[1]{Variable Energy Cyclotron Centre, Kolkata 700064, India.}
	\affil[2]{Homi Bhabha National Institute, Training School Complex, Anushaktinagar, Mumbai - 400094, India}
	
	\abstract{
		We study the spectral properties of a multiparametric system 
		having particle-hole symmetry in random matrix setting. We observe 
		a crossover 
		from Poisson to Wigner-Dyson 
		like behavior in average local
		ratio of spacing within a spectrum of 
		single matrix as a function of effective single parameter referred 
		to as 
		complexity parameter. The average local
		ratio of spacing varies 
		logarithmically in complexity parameter across the transition. 
		This 
		behavior is universal for different ensembles subjected to same 
		matrix 
		constraint like particle-hole symmetry. The universality of this 
		dependence is further established by studying interpolating 
		ensemble 
		connecting systems with particle-hole symmetry to that with chiral 
		symmetry. For each interpolating ensemble the behavior remains 
		logarithmic 
		in complexity parameter. We verify this universality of 
		spectral fluctuation  in case of a 2D  Su-Schrieffer-Heeger (SSH) 
		like 
		model along with the logarithmic dependence on complexity 
		parameter for
		ratio of spacing during transition from 
		integrable to non-integrable limit.}
	
	\vspace{2pc}

	\maketitle

	\section{Introduction}
	Statistical behavior of a complex system modeled by a random matrix 
	ensemble (RME) requires information of different constraints imposed on 
	the Hamiltonian matrix representing the system  \cite{haake2010, mehta91, 
		shukla2012generalized, guhr1998random}. The constraints influencing the 
	nature of a system-dependent RME are categorized in two broad types: a) 
	\textit{matrix or global constraints} which mainly affect broad structure 
	of single matrix governing the system; b) \textit{ensemble or local 
		constraints} which specify the distribution properties of the matrix 
	elements \cite{haake2010, mehta91}. The symmetries are one of the examples 
	of matrix constraints whereas disorder in the matrix elements is an 
	example of ensemble constraints \cite{haake2010, mehta91, 
		shukla2012generalized}. Matrix constraints time-reversal symmetry and 
	spin-rotation symmetry together lead the RME to Gaussian orthogonal 
	ensemble (GOE) as it becomes invariant under orthogonal transformation 
	\cite{haake2010, mehta91}. Absence of both or one of these symmetries 
	leads to the other two ensembles (Gaussian unitary ensemble (GUE) and 
	Gaussian symplectic ensemble (GSE), respectively) of Dyson's 
	three-fold way of canonical 
	transformations. Beyond this classification, three chiral symmetric and 
	four particle-hole symmetric ensembles were introduced later along with or 
	without time-reversal and spin-rotation symmetry 
	\cite{altland1997nonstandard}. 
	The choice of independence of matrix elements leads to 
	Gaussianity of the distribution, and further specification of ensemble 
	constraints 
	like mean and variance of 
	the disorder distribution of the matrix elements lead to 
	different kind of ensembles. Alternatively, one could add the 
	correlation in the matrix elements to achieve the same
	\cite{shukla2012generalized, 
		mondal2020spectral, shukla2005level, shukla2000alternative, 
		shukla2005random}.
	Variation of ensemble parameters alone has been 
	shown to drive the system from Poissonian limit to Wigner-Dyson statistics 
	(\textit{e.g.} metal-insulator transition in Anderson ensemble 
	\cite{evers2008anderson}). 
	Moreover, irrespective of ensemble parameters, the local 
	spectral statistics for different ensembles belonging to 
	same matrix constraint group show similar behavior based on the 
	\textit{ensemble 
		complexity 
		parameter}\cite{shukla2000alternative,shukla2005level,shukla2005random,
		dutta2007complex,shukla2008towards,mondal2020spectral}. This is a single 
	parameter representing the degree and nature of complexity in the system defined as a scaled logarithmic function of all parameters of the 
	ensemble. This complexity parameter based universality 
	of spectral fluctuations for different ensembles with and without chiral 
	symmetry for Dyson's Gaussian ensembles were studied in 
	\cite{mondal2020spectral} and \cite{shukla2005level, 
		shukla2000alternative, dutta2007complex} respectively based on the 
	complexity parameter.  It is natural to ask if this type of universality
	remains preserved for different ensembles with matrix constraint 
	particle-hole symmetry, completing the additional symmetry classes 
	considered in \cite{altland1997nonstandard}. This symmetry naturally 
	occurs in various physical systems, one of the well known cases is 
	Andreev reflection at the interface of the normal metal-superconducting 
	systems 
	leaving the associated Hamiltonian invariant under this 
	symmetry \cite{andreev1964thermal}. 
	The standard Hamiltonian used for these systems is the 
	Bardeen-Cooper-Schrieffer (BCS) Hamiltonian which can be expressed in 
	terms of another Hermitian operator, known as Bogoliubov deGennes (BdG) 
	Hamiltonian (Eq.~\ref{1}) \cite{andreev1964thermal, 
		shaginyan2007asymmetric, cooper1956bound, fernandes2020lecture}.
	The Su-Schrieffer-Heeger (SSH) model is a topological example of 
	particle-hole symmetric Hamiltonian. The one dimensional representation of 
	this model is popular to investigate topological phenomena in condensed 
	matter physics \cite{zirnbauer2021particle, asboth2016short, 
		su1979solitons}. Later, people generalized this model to several types of 
	two dimensional (2D) SSH model to study different properties of topology 
	\cite{heeger1988solitons, ota2018topological, ota2019photonic, 
		xie2018second, xie2019visualization, chen2018two, chen2019direct, 
		zheng2019observation, kim2020topological, li2021tunable, 
		li2022topological}. For further appearances of particle-hole symmetry in 
	condensed matter physics, see reference 
	\cite{zirnbauer2021particle}.	
	Considering the Hamiltonian matrix with particle-hole symmetry described 
	in the site basis, the off-diagonal blocks of it represent the hopping of 
	the particles (electrons or holes) from one site to others while the onsite dynamics of those particles are represented by its 
	diagonal blocks  \cite{altland1997nonstandard, haake2010, 
		simons2000lecture, gnutzmann2004universal-1, gnutzmann2004universal-2}. 
	The latter ones are considered to be zero if the system has chiral 
	symmetry instead of particle-hole symmetry \cite{haake2010, 
		verbaarschot1994spectrum, gnutzmann2004universal-1, 
		gnutzmann2004universal-2}. A transition from particle-hole to chiral 
	symmetry can be achieved by controlling the ensemble 
	parameters of the diagonal 
	blocks of Hamiltonian matrix. 
	Motivated by this, we ask 
	two specific but related questions: a) Whether transition from Poissonian 
	to Wigner-Dyson like behavior within single spectrum is governed by the 
	complexity parameter and show universality for different ensembles respecting the particle-hole symmetry as matrix constraint? and b) Whether each point of transition from particle-hole to chiral symmetric ensembles respects this universality?

	In this paper we consider time-reversal symmetric systems with integer 
	spin having particle-hole symmetry. In section \ref{s2}, we introduce the 
	matrix representation of the Hamiltonian and discuss the relevant 
	symmetries along with the ensemble density.
	We formulate single parametric representation of multiparametric 
	evolution of matrix elements in section \ref{s3} following the steps 
	for other symmetry classes discussed in \cite{mondal2020spectral, 
	shukla2005level, shukla2000alternative, shukla2005random}. 
	Section \ref{s5} details three different ensembles arising 
	due to different ensemble constraint even though they satisfy the same 
	matrix constraint.  In section \ref{s6}, we present our studies on 
	spectral statistics across the spectrum and show that Poisson to 
	Wigner-Dyson transition is independent of the ensembles considered. We propose an interpolating ensemble from particle-hole to chiral symmetry in section \ref{s7} and study the ratio of nearest neighbor spacing distribution. We show that at each interpolating point, the transition across the spectrum remains identical for different ensembles. 
	These findings are further corroborated by studying a dynamical model of 
	2D SSH type in section \ref{s9}. The summary and outlook follows in 
	section \ref{s10}.
	
	\section{Multiparametric Gaussian ensemble with particle-hole symmetry and other matrix constraints}
	\label{s2}
	A $2N\times 2N$ Hamiltonian matrix with particle-hole symmetry (also known as BdG Hamiltonian) can be described as
	\begin{equation}
	H_{ph}=
	\begin{pmatrix}
	\mathcal{H} & \Delta\\
	\Delta^{\dagger} & -\mathcal{H} ^T
	\end{pmatrix}
	\label{1}
	\end{equation}
	with $\mathcal{H}$ is a $N\times N$ hermitian block matrix: 
	$\mathcal{H}=\mathcal{H}^\dagger$ and $\Delta$ is a $N\times N$ 
	anti-symmetric block matrix, \textit{i.e.} $\Delta= -\Delta^T$ due to 
	Fermi statistics \cite{altland1997nonstandard, haake2010, mehta91}.
	
	The particle-hole symmetry exchanges electrons with holes. Considering an 
	antiunitary operator $\mathcal{P}=\sigma_x\mathcal{K}$, where 
	$\mathcal{K}$ is the complex conjugation operator and $\sigma_x$ is the 
	Pauli matrix acting on the blocks. Therefore, the Hamiltonian in Eq.~\ref{1} can be described as
		\begin{equation}
	H_{ph}=-\mathcal{P}H_{ph}\mathcal{P}^{-1}.
	\end{equation}
	The minus sign due to the particle-hole symmetry implies the spectrum of $H_{ph}$ must be symmetric around zero energy and for every eigenvector $\psi$ of $H_{ph}$ with energy $E$, there will be a particle-hole symmetric eigenvector $\mathcal{P}\psi$ with energy $-E$.
	
	If time-reversal symmetry along with spin-rotation symmetry is also 
	present in the system with particle-hole symmetry, the Hamiltonian 
	$H_{ph}$ become invariant under orthogonal transformation and hence, 
	the matrix elements become real. 
	Now both the blocks $\mathcal{H}$ and $\Delta$ become real 
	\begin{align}
	\mathcal{H}= & \mathcal{H}^* = \mathcal{H}^T\\
	\Delta &= \Delta^T.
	\end{align}
	Therefore, the Hamiltonian $H_{ph}$ in Eq.~\ref{1} will look like
	\begin{equation}
	H=
	\begin{pmatrix}
	\mathcal{H} & \Delta\\
	\Delta & -\mathcal{H}
	\end{pmatrix}.
	\label{5}
	\end{equation}
	This is called class CI of particle-hole symmetry. In this paper, we 
	confine our study only to this CI class. The number of independent 
	elements in $H$ now reduced to $N(N+1)$ from $4N^2$. The results we have 
	achieved in this paper are intuitively true for the other three classes of 
	particle-hole symmetry in qualitative manner; although the technical difficulties in the calculation would increase.

		The joint probability distribution function of the ensemble of such matrices is given by
		\begin{align}
		\rho(H)= \rho_{\mathcal{H}}(\{H_{k,l}\})\rho_{\Delta}(\{H_{k,l+N}\})
		\label{eq6}
		\end{align}
		with $k,l$ going from $1\to N$. The additional constraints of 
		sum of diagonal blocks and difference of off-diagonal blocks 
		giving null matrices have been taken into account to write 
		Eq.~\ref{5}.
		$\rho_{\mathcal{H}}(\mathcal{H})$ and $\rho_{\Delta}(\Delta)$ in Eq.~\ref{eq6} are	the probability densities of the ensemble of real symmetric $\mathcal{H}$ and $\Delta$ matrices respectively, defined as
		\begin{align}
		\rho_{\mathcal{H}}(\mathcal{H},h,b)=\mathcal{C_H} \;{\rm exp}\Bigg[-\sum_{k\leq l =1}^N \frac{1}{2h_{kl}^{(\mathcal{H})}}(\mathcal{H}_{kl}-b_{kl}^{(\mathcal{H})})^2\Bigg]
		\label{rho1}\\
		\rho_{\Delta}(\Delta,h,b)=\mathcal{C}_{\Delta} \;{\rm exp}\Bigg[-\sum_{k\leq l =1}^N 
		\frac{1}{2h_{kl}^{(\Delta)}}(\Delta_{kl}-b_{kl}^{(\Delta)})^2\Bigg]
		\label{rho2}
		\end{align}
		where $\mathcal{C_H}$ and $\mathcal{C}_{\Delta}$ are normalization constants. 
		All the matrix elements are independent and identically distributed 
		and taken randomly from a Gaussian distribution with variances $h_{kl}^{(\mathcal{H})}$ and $h_{kl}^{(\Delta)}$ and means $b_{kl}^{(\mathcal{H})}$ and $b_{kl}^{(\Delta)}$. 
		Using $H_{kl}=\mathcal{H}_{kl}$ and $H_{k,N+l}=\Delta_{kl}$, the ensemble density of $H$ becomes
		\begin{align}
		\rho(H,h,b)= \mathcal{C} \;{\rm exp}\Bigg[-\sum_{k\leq l =1}^N \frac{1}{2h_{kl}}(H_{kl}-b_{kl})^2
		-\sum_{k\leq l =1}^N 
		\frac{1}{2h_{k,N+l}}(H_{k,N+l}-b_{k,N+l})^2\Bigg],
		\label{6}
		\end{align}

	Let us define $h\equiv[h_{kl}]$ and $b\equiv[b_{kl}]$ to be the matrices 
	for variances and means respectively. Clearly,  
	with different choices of ensemble constraints \textit{i.e.} $h$ and $b$, 
	it is possible to achieve different ensembles with particle-hole symmetry. 
	We discuss three of them later in this paper.
	
	\section{Diffusion of matrix elements: Ensemble complexity parameter}
	\label{s3}
	Considering our system (Eq.~\ref{5}) evolves in the block matrix space by 
	a continuous variation of multiple parameters ($h_{kl}$ and $b_{kl}$ in 
	our case), preserving the global symmetries in a suitable basis, we can 
	use previously studied idea of the diffusion of the density 
	depending on the multiple 
	parameters via a single averaged logarithmic function $ Y $ 
	referred to as the ensemble complexity parameter \cite{mondal2020spectral, shukla2005level, shukla2000alternative, shukla2005random}. 	
	The diffusion equation connects evolution in parametric space to that in matrix space.
	$ \mathcal{H} $ and $ \Delta $ themselves being Hermitian, 
	following \cite{shukla2005level}, single parametric diffusion equation 
	of 	$\rho_{\mathcal{H}}(\mathcal{H})$ and $\rho_{\Delta}(\Delta)$ 
	in matrix space is given by,
		\begin{align}
		\frac{\partial \rho_{\mathcal{H}}}{\partial Y_{\mathcal{H}}} =  \sum_{k,l} \frac{\partial}{\partial \mathcal{H}_{kl}}\left[\left(\frac{1+\delta_{kl}}{2}\right) \frac{\partial  \rho_{\mathcal{H}}}{\partial \mathcal{H}_{kl}}+\gamma \; \mathcal{H}_{kl} \; \rho_{\mathcal{H}}\right]\\
		\frac{\partial \rho_{\Delta}}{\partial Y_{\Delta}} =  \sum_{k,l} \frac{\partial}{\partial\Delta_{kl}}\left[\left(\frac{1+\delta_{kl}}{2}\right) \frac{\partial  \rho_{\Delta}}{\partial \Delta_{kl}}+\gamma \; \Delta_{kl} \; \rho_{\Delta}\right]
		\label{ch8}	
		\end{align}
		where the complexity parameters $Y_\mathcal{H}$ and $ Y_\Delta $ 
		are,
		\begin{align}
		Y_{\mathcal{H}}= -\frac{1}{2M_{\mathcal{H}}\gamma} {\rm ln} \Big[\prod_{k\leq l = 1}^N|1-\gamma \;g_{kl}\;h_{kl}^{(\mathcal{H})}|\;|b_{kl}^{(\mathcal{H})}|^2\Big]+C_{\mathcal{H}}\\
		Y_{\Delta}= -\frac{1}{2M_{\Delta}\gamma} {\rm ln} \Big[\prod_{k\leq l = 1}^N|1-\gamma \;g_{kl}\;h_{kl}^{(\Delta)}|\;|b_{kl}^{(\Delta)}|^2\Big]+C_{\Delta}
		\label{eqy}
		\end{align} 
		with $g_{kl}=2-\delta_{kl}$ while $\gamma$ is taken as arbitrary 
		parameter.
		Here $M_{\mathcal{H}}=M_{\Delta}\equiv M$ (since size, symmetry and matrix elements distribution of $\mathcal{H}$ and $\Delta$ are same.) is the number of non-zero $h_{kl}$ and $b_{kl}$ and the $\prod$s are 
		over them. $C_{\mathcal{H}}$ and $C_{\Delta}$ are the integration constants.	 
		Therefore, the solution of complexity parameter for $\rho(H)$ (Eq.~\ref{6}) is 
		\begin{align}
		Y= -\frac{1}{2M\gamma} \Big[{\rm ln} \Big\{\prod_{k\leq l = 1}^N|1-\gamma \;g_{kl}\;h_{kl}|\; |b_{kl}|^2\Big\}
		+ {\rm ln} \Big\{ \prod_{k\leq l = 1}^N |1-\gamma \;g_{kl}\;h_{k,N+l}| \;|b_{k,N+l}|^2\Big\} \Big]\nonumber\\ + \;{\rm const.}
		\label{8}
		\end{align}
		%
		

	It is expected that diffusion of density would lead to an evolution in the 
	spectral properties of the systems. We next explore this numerically for 
	the system with particle-hole symmetry. 
	The spectral fluctuation is dependent on the ensemble constraints and due to non-stationarity of the spectrum, it is 
		necessary to rescale the complexity parameter $Y$ by the square of 
		local mean level spacing $\Delta_l(e)$ at the spectral point of 
		interest (say e). The re-scaled complexity parameter $\Lambda$ can 
		be written as
%
	\begin{equation}
	\Lambda(e,Y)= \frac{Y-Y_0}{\Delta_l^2(e)}
	\label{9}
	\end{equation}
	with $Y_0$ being the complexity parameter of initial state
	 at the start of evolution. Since the eigenstates are not always 
	expected to be exactly delocalized, $\Delta_l(e)$ strongly depends on 
	energy range. It corresponds to states which are interacting at energy 
	$e$ only. As we know, localized eigenstates in different	parts of 
	the basis space do not interact, we expect $\Delta_l(e)$ is 
	proportional to average correlation
		at energy e. Therefore, an acceptable definition of this can be
	\begin{equation}\label{eq:rho_local}
	\Delta_l(e)\equiv\frac{1}{\langle\rho_{loc}(e)\rangle}, \quad 
	\text{with } \;\rho_{loc}(e)\equiv\sum_n\phi_n\delta(e-e_n)
	\end{equation}
	where $\phi_n$ is the $n^{\rm th}$ eigenfunction overlapping with other 
	eigenfunctions in the energy range $e$ and $ e\pm\delta e $ 
	\cite{mondal2020spectral}. The sum in Eq. \ref{eq:rho_local}
	is over only the overlapping eigenfunctions (locally) around $e$ 
	and the $\langle.\rangle$ is also a local average. It is to be noted that 
	in case of the global level density: 
	$R_1(e)=2N\langle\rho(e)\rangle=\sum_k\langle\delta(e-e_k)\rangle$, the 
	sum is over entire energy range and the $\langle.\rangle$ over $\delta$ is 
	local around energy $e$.
	Therefore,
	\begin{equation}
	\langle\rho_{loc}(e)\rangle=\frac{\zeta}{2N}R_1(e)
	\end{equation}
	where $\zeta$ is the correlation volume of eigenstates and $2N$ is the 
	size of the 
	Hamiltonian matrix $H$. The range of the value of $\zeta$ is as follows: 
	for delocalized eigenstates $\zeta=2N$ and for extremely localized case 
	$\zeta=1$ \footnote{Note that, here one can expect $\zeta$ to be equal 
		to the participation ratio. (\textit{i.e.} $\zeta=\frac{1}{\langle I_2(e)\rangle}$), but 
		this is not true if the inverse participation ratio (IPR) $I_2(e)$ varies 
		with energy $e$.}. Therefore, $\langle \rho_{loc}(e)\rangle$ changes from 
	$\frac{1}{2N}R_1(e)$ to $R_1(e)$ as we go from localized to delocalized 
	eigenstates, \textit{i.e.} $\frac{1}{2N}R_1(e)<\langle \rho_{loc}(e)\rangle <R_1(e)$. Eq.~\ref{9} can be written as
	\begin{equation}
	\Lambda(e,Y)=(Y-Y_0) \Big(\frac{\zeta R_1(e)}{2N}\Big)^2.
	\label{33}
	\end{equation}
	
	This rescaled complexity parameter not only have the information about 
	different parameters (mean and variance in case of Gaussian distribution) 
	of 
	the distribution but also depends on density and the localization 
	properties of the spectrum at the point of interest $e$.
	
	\section{Details of different ensembles with  same matrix constraints}
	\label{s5}
	In this section, we present the details of the different ensembles arising 
	due to different ensemble constraints despite preserving same matrix 
	constraint. Depending on the choice of the form of variance in Eq.~\ref{6} 
	which serves 
	the purpose of ensemble constraint here, we study three ensembles (a) 
	Brownian (b) Anderson and (c) exponential ensemble. All these three 
	ensembles have the particle-hole, time-reversal and spin rotation 
	symmetry. 
	
	\subsection{Brownian ensemble (BE)}
	\label{s5a}
	\new{The evolution of a stationary ensemble under  perturbation from 
	one stationary limit to another can be described by Brownian motion 
	model \cite{dyson1962brownian}. The non-stationary intermediate states 
	of diffusion connecting two equilibrium ensembles at both ends can be 
	represented by a Brownian ensemble 
		 \cite{shukla2005level}. The initial integrable individual 
		 blocks of $H$ in Eq.~\ref{5} under the basis-independent Gaussian 
		 perturbation can transit to chaotic limit preserving all the 
		 symmetries imposed on $H$.		
}
	The variance and mean of the \new{matrix elements distribution} in Eq.~\ref{6} are 
	\new{chosen to be}
	\begin{align}
	h_{kk}=h_{k,N+k}=\frac{1}{2\gamma}
	\label{19}\\
	h_{kl}=h_{k,N+l}=\frac{h_{kk}}{2(1+\mu)}
	\label{20}\\
	b_{kl}=b_{k,N+l}=0 \qquad \forall \; k,l.
	\label{21}
	\end{align}
	Here $\mu\equiv cN^2$ with $N$ as the dimension of $\mathcal{H}$ (which is 
	also the dimension of $ \Delta $)  and $c$ is an arbitrary parameter. 
	\new{Zero-mean of the Gaussian random variable ensures each block of $H$ to 
	be diagonal in $c\to \infty$ limit, while renders GOE behavior in $c\to 0$ limit.
%
%
%
%
   }The ensemble density (Eq.~\ref{6}) in these parameters takes the form,
	\begin{align}
	\rho(H,h,b)&= \mathcal{C} \;{\rm exp}\Bigg[-\gamma \sum_k H_{kk}^2 -2\gamma(1+\mu)\sum_{k<l =1}^N |H_{kl}|^2 \nonumber\\
	&-\gamma\sum_k H_{k,N+k}^2 -2\gamma(1+\mu)\sum_{k<l =1}^N |H_{k,N+l}|^2 
	\Bigg]\mathcal{F}_s,
	\end{align}
	while the complexity parameter (Eq.~\ref{8}) will become
	\begin{align}
	Y= \frac{N}{M\gamma}\Big[{\rm ln}(2) -\frac{N-1}{2}\;{\rm 
		ln}\Big\{1-\frac{1}{2(1+\mu)}\Big\}\Big].
	\end{align}
	For the initial density when we have zero hopping, the complexity 
	parameter $Y$ 
	is $Y_0=\frac{N}{M\gamma}\;{\rm ln}(2)$. 
	Subtracting this initial $ Y_0 $ from $ Y $ and putting the number of 
	non-zero independent elements $M=N(N+1)/2$, 
	\begin{align}
	Y-Y_0 =-\frac{(N-1)}{\gamma (N+1)}\;{\rm 
		ln}\Big\{1-\frac{1}{2(1+\mu)}\Big\}.
	\end{align}
	Expanding the logarithm (ignoring the higher order terms), in the limit 
	$N>>1, \mu>>1$, we obtain
	\begin{align}
	Y-Y_0\simeq \frac{1}{2\gamma \mu} .
	\label{eq:compl_para_BE_y-y0}
	\end{align}
	
	In case of Brownian ensemble, the eigenfunctions are delocalized over entire Hilbert space \cite{shukla2005level, mondal2020spectral} leading the correlation volume $\zeta$ in Eq.~\ref{33} to be equal to the matrix size $2N$. Therefore, the rescaled complexity parameter $\Lambda$ in this case is given by
	\begin{equation}
	\Lambda_B(e,Y)=(Y-Y_0)R_1(e)^2
	\label{28a}
	\end{equation}
	with $Y-Y_0$ provided in Eq.~\ref{eq:compl_para_BE_y-y0}. Throughout this 
	paper, we have considered the numerical value of the arbitrary parameter 
	$\gamma$ in the definition of $Y$ in Eq.~\ref{8} to be equal to $1/4$. Now 
	substituting the values of $\gamma$ and $\mu$, we have
	\begin{equation}
	\Lambda_B(e)=\frac{2}{cN^2}R_1(e)^2.
	\label{36}
	\end{equation}
	In all the calculation, $ c $ is the true free parameter for this ensemble.
	\subsection{Anderson Ensemble (AE)}
	\label{s5b}
	\new{Transport properties of a conduction band electron in presence of 
	impurities was first studied by Anderson by writing down a 
	tight-binding model with nearest neighbor interaction 
	\cite{anderson1958absence}. The ensemble of such random Hamiltonians  
	are often referred as Anderson ensemble \cite{shukla2005level}.
%
 The dimension of the system decides the number of nearest neighbors and accordingly determines the sparsity of the Hamiltonian matrix in site basis. Here, the individual blocks in $H$ in Eq.~\ref{5} represent 2D Anderson ensemble and the variances and mean of their elements in Eq.~\ref{6} are taken as } 
	\begin{align}
	h_{kk}=h_{k,N+k}=\frac{w^2}{12},&\qquad b_{kk}=0
	\label{h}\\
	h_{kl}=h_{k,N+l}=f_1(k,l)\frac{w_s^2}{12},&\qquad b_{kl}=b_{k,N+l}=f_2(k,l)t_s
	\label{b}
	\end{align}
	with $f_1(k,l)$ and $f_2(k,l)$ are equals to $1$ for connected sites on a 
	two dimensional lattice and otherwise zero, $w$ and $w_s$ are arbitrary 
	parameters. Here we consider $t_s=0$, so that mean of the \new{hopping elements distribution} becomes zero. \new{All the mean of the distribution are taken to be zero, so that the sparsity of the Hamiltonian matrix can be controlled by variances only. Keeping $w_s$ fixed and varying $w$ from $\sim w_s$ to higher value leads the eigenstates of $H$ from delocalized to localized ones.} The ensemble density (Eq.~\ref{6}) in this case is,\\

	\begin{align}
	\rho(H,h,b) &= \mathcal{C} \;{\rm exp}\Bigg[-\frac{6}{w^2}\sum_{k=1}^N H_{kk}^2 -\frac{6}{w_s^2}\sum_{k<l =1}^N \frac{1}{f_1(k,l)}|H_{kl}|^2 \nonumber\\
	& -\frac{6}{w^2}\sum_{k=1}^N H_{k,N+k}^2 -\frac{6}{w_s^2}\sum_{k<l =1}^N 
	\frac{1}{f_1(k,l)}|H_{k,N+l}|^2 \Bigg]\mathcal{F}_s,
	\end{align}
	while the complexity parameter from Eq.~\ref{8} becomes
	\begin{align}
	Y= -\frac{N}{M\gamma}\Big[{\rm ln} \Big 
	|1-\gamma\frac{w^2}{12}\Big|+\frac{z}{2}\;{\rm ln}\Big 
	|1-\gamma\frac{w_s^2}{6}\Big | \Big]+Y_0,
	\label{27}
	\end{align}
	where $z$ is the number of nearest neighbor (non-zero) elements for each 
	row of the matrix. $Y_0$ is the initial complexity parameter. 
	
	For all the numerical work, we consider two dimensional Anderson 
	Hamiltonian \textit{i.e.} $z=4$. We have chosen the 
	variances of the off-diagonal elements of block matrices as
	$w_s^2=12$ which fixes the hopping parameter. Substituting these 
	values along with that of $\gamma 
	(=1/4)$, we get 
	\begin{align}
	Y-Y_0= -\frac{4}{N+1}\;{\rm ln} \Big |1-\frac{w^2}{48}\Big|+\frac{8}{N+1}\;{\rm ln}(2)
	\label{43}
	\end{align}
	with the number of non-zero independent elements $M=N(N+1)$. Here the free parameter is $w^2$.

	The correlation volume $\zeta$ in Eq.~\ref{33} is taken  intuitively 
	as participation ratio, \textit{i.e.} $\zeta=\frac{1}{\langle 
	I_2(e)\rangle}$, with $\langle.\rangle$ to be ensemble average of the 
	local average around 
	energy $e$. The re-scaled complexity parameter in this case is given by 
	\begin{equation}
	\Lambda_A(e,Y)=(Y-Y_0) \Bigg(\frac{R_1(e)}{2N\langle I_2(e)\rangle}\Bigg)^2
	\label{44a}
	\end{equation}
	where $(Y-Y_0)$ is provided in Eq.~\ref{43}.

	\subsection{Ensemble with exponential decay (EE)}
	\label{s5c}
	\new{A random band matrix model when no longer bounded by nearest 
	neighbor interactions becomes less sparse. Considering the hopping is 
	random and the variances of this Gaussian random hopping decay with 
	the increasing distances between two adjacent lattice sites, one of 
	the possible ways to model such system is exponential ensemble 
	(introduced in \cite{fyodorov1991scaling}).		
	Here, the block matrices of $H$ in Eq.~\ref{5} belong to exponential ensemble where the variances of their matrix elements distribution show exponential decay from their diagonals \cite{mondal2020spectral}. The exponential suppressed variances in Eq.~\ref{6}}
are given by
	\begin{equation}
	h_{kl}=h_{k,N+l}=\frac{1}{{\rm exp}\big(\frac{|k-l|}{\tilde{b}}\big)^2}
	\label{45}
	\end{equation}
	where $\tilde{b}$ is an arbitrary parameter. \new{As $\tilde{b}\to \infty$, the Hamiltonian becomes completely chaotic whereas decreasing finite value of $\tilde{b}$, localizes the system. At initial state when no hopping is introduced, \textit{i.e.} at $\tilde{b}\to 0$, individual blocks of $H$ are integrable since} the mean of the distribution is chosen to be  $b_{kl}=b_{k,N+l}=0$ for all $k$ and $l$. \new{The sparsity as well as the structure of the Hamiltonian matrix $H$ can be determined by tuning only the parameter $\tilde{b}$.} The 
	probability distribution (Eq.~\ref{6}) of this ensemble becomes
	\begin{align}
	\rho(H,h,b) &= \mathcal{C} \;{\rm exp}\Bigg[-\frac{1}{2}\sum_k H_{kk}^2 -\sum_{k<l =1}^N \frac{{\rm exp}\big(\frac{|k-l|}{\tilde{b}}\big)^2}{2} |H_{kl}|^2 \nonumber\\
	& -\frac{1}{2}\sum_k H_{k,N+k}^2 -\sum_{k<l =1}^N \frac{{\rm exp}\big(\frac{|k-l|}{\tilde{b}}\big)^2}{2} |H_{k,N+l}|^2 \Bigg]\mathcal{F}_s.
	\label{39}
	\end{align}
	The complexity parameter $Y$ from Eq.~\ref{8}
	takes the form
	\begin{align}
	Y= -\frac{1}{M\gamma}\Big[N\;{\rm ln}(1-\gamma) +\sum_{r=1}^{N-1}(N-r){\rm ln} \Big|1-\frac{2\gamma}{{\rm exp}\big(\frac{r}{\tilde{b}}\big)^2}\Big|\Big]
	\end{align}
	with $r\equiv|k-l|$. Subtracting initial state complexity 
	$Y_0=-\frac{N}{M\gamma}{\rm ln}(1-\gamma)$ yields 
	\begin{equation}
	Y-Y_0=-\frac{8}{N(N+1)}\sum_{r=1}^{N-1}(N-r){\rm ln} \Big|1-\frac{1}{2\;{\rm exp}(\frac{r}{\tilde{b}})^2}\Big|
	\label{47}
	\end{equation}
	with the number of non-zero independent matrix elements $M$ as equal to 
	$\frac{N(N+1)}{2}$ and $\gamma=1/4$. The free parameter is $\tilde{b}$.

	In this ensemble, the correlation volume $\zeta$ in Eq.~\ref{33}  is  
	considered as $\zeta=\frac{1}{N^{1/4}\langle I_2(e)\rangle}$, the 
	re-scaled complexity parameter $\Lambda$ is now given by
	\begin{equation}
	\Lambda_E(e,Y)=(Y-Y_0)\Bigg(\frac{R_1(e)}{2N\times N^{1/4}\langle I_2(e)\rangle}\Bigg)^2
	\label{49}
	\end{equation}
	with $(Y-Y_0)$ described as in Eq.~\ref{47}.

	\section{Transition from Poisson to Wigner-Dyson statistics in single spectrum: universality}
	\label{s6}
	Coming to first of the two questions we posed in introduction about 
	the behavior of spectral statistics as a function of energy, we choose 
	to study short range correlations, namely, ratio of nearest neighbor 
	spacing distribution. First, to understand the spectral properties of 
		particle-hole symmetric ensembles, let's recall the discussion in Ref. 
		\cite{altland1997nonstandard} briefly. The joint probability 
		distribution function of eigenvalues of the CI variant of 
		particle-hole symmetric Hamiltonian is given by,
		\begin{equation}
		P(\{e\})\;d\{e\}\propto\prod_{i<j}^{2N}|e_i^2-e_j^2| 
		\prod_{k}^{2N}|e_k|\; {\rm exp}^{-e_k^2/h^2} de_k
		\label{p}
		\end{equation}
		where $|e_k|$ represents the absence of energy eigenvalues at  
		origin which is attributed to the  
		interaction of $k$th level $e_k$ with its ``image" $-e_k$. 
		The term $|e_i^2-e_j^2|$ describes mutual repulsion between levels and 
		the factor $(e_i+e_j)$ in $|e_i^2-e_j^2|$ is due to the interaction of 
		$e_i$th level with the ``image" of $e_j$th level.
		Now, to study the spectral fluctuation in terms of nearest neighbor spacing ratio, we consider the symmetry reduced spectrum by choosing 
		positive eigenlevels only. And in the absence of ``image'' of $ e 
		$'s, according to above discussion the relevant spacing 
		distribution will be Wigner-Dyson (see Appendix \ref{nnspacing} for an 
		illustrative calculation).

		The numerical diagonalization of matrices are done using  LAPACK (a standard software library for numerical linear algebra subroutines for complex matrices \footnote{The LAPACK subroutines used for diagonalization of complex matrices are available at \url{https://www.netlib.org/lapack}}) subroutines. The numerical parameters chosen for this study is detailed as follows. 
	Throughout our analysis, we have considered $N\times N$ block matrices for 
	$2N\times 2N$ Hamiltonian $H$. The mean ($\{b_{kl}\}$) of the Gaussian 
	distribution for all the cases is considered to be zero to keep our 
	numerical study simple. The numerical value of the arbitrary parameter 
	$\gamma$ in the definition of $Y$ in Eq.~\ref{8} is fixed to be equal to 
	$1/4$. Ensemble specific details are: (\textbf{BE}) This ensemble of 
	matrices are 
	exactly diagonalized for two $c$ values ($c=0.4$ and $1$), while $ 2N = 
	1024 $ and the size of the ensemble is taken 5000. (\textbf{AE}) Two 
	dimensional 
	(2D) Anderson ensemble is considered for each block matrices 
	($\mathcal{H}$ and $\Delta$) of the Hamiltonian in Eq.~\ref{5} with linear 
	size $L=22$ and therefore, the size of $H$ becomes $2N=2\times L^2=968$. 
	An ensemble of $5000$ matrices of this size is considered to diagonalize 
	for each cases with $w^2=12$ and $36$ in the variance of the diagonal 
	elements of the block matrices in Eq.~\ref{h}. For all the cases the 
	variance of the off-diagonal elements are fixed at $w_s^2=12$ and the mean 
	of the distribution are kept to be zero (\textit{i.e.} $t=t_s=0$). (\textbf{EE}) 
	We have considered two different cases with $\tilde{b}=10$ and $12$ and 
	generated $5000$ matrices of size $2N=1024$ for both the cases for 
	exact  diagonalization.

\begin{figure}[ht!]
		\centering
		\includegraphics[width=0.38\textwidth]{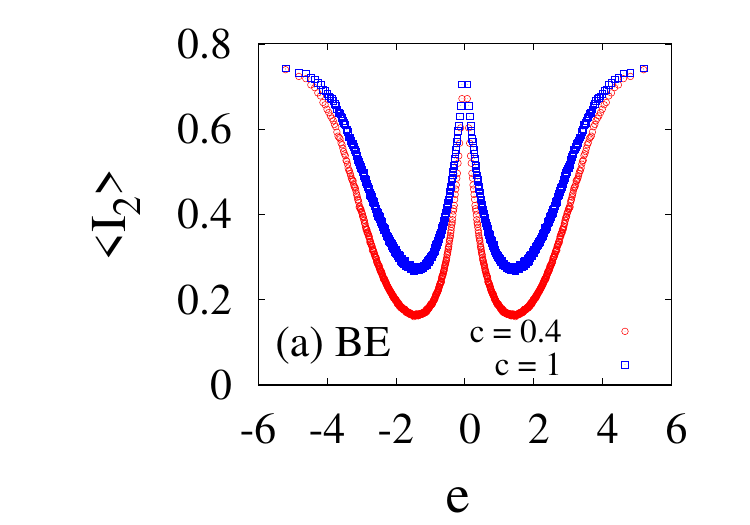}
		\hspace{-1.2cm}
		\includegraphics[width=0.38\textwidth]{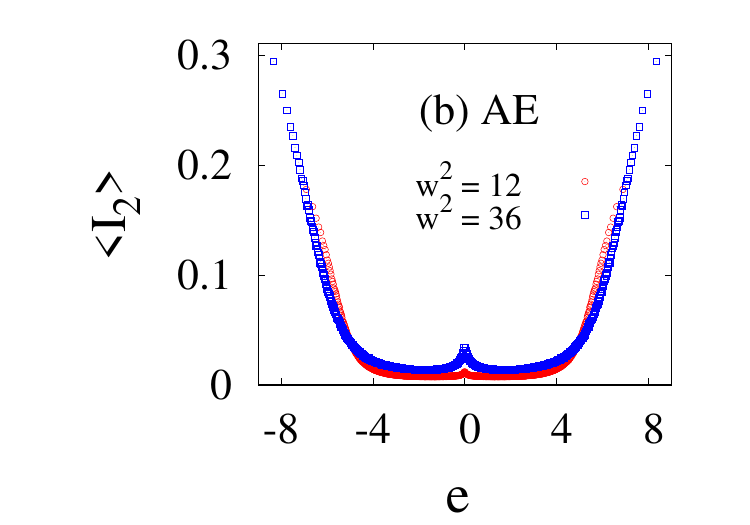}
		\hspace{-1.2cm}
		\includegraphics[width=0.38\textwidth]{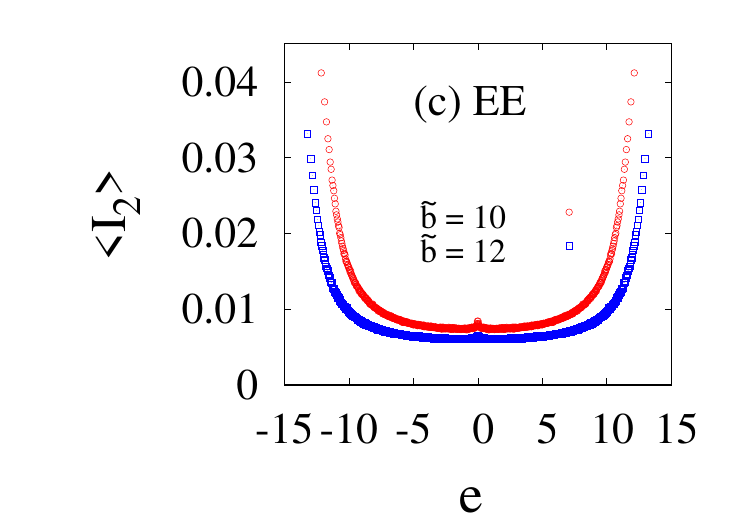}
		\caption{$\langle I_2\rangle$ vs $e$ for (a) BE, (b) AE and (c) EE for two different variances at each case. The $\langle I_2\rangle$ is sensitive to the variances of the individual ensembles as well as the type of the ensembles.}
		\label{fig2}
	\end{figure}

The effect of variances, \textit{i.e.} ensemble constraint on the spectrum despite coming from same matrix constraint when looked at as a function of energy is observed by studying the density of states (see the details in Appendix \ref{DOS}).
To explore this ensemble constraint dependence further, we look at the 
	energy dependence of eigenstates properties as well like inverse 
	participation ratio (IPR).	
	The IPR is defined as the inverse of the number of basis states 
	participating in the wavefunction. The inverse participation ratio for the 
	$n^{th}$ eigenstate $\psi_n$ corresponding to 
	the eigenvalue 	$e_n$, can be expressed as 
	$I_2(\psi_n)=\sum_{k=1}^{2N} \lvert \psi_{kn}\rvert^4$ which is
	second order moment of local intensity \cite{haake2010, mehta91, 
		brody1981random}. 	
	Figure~\ref{fig2} shows the variation of ensemble averaged IPR, $\langle 
	I_2(e)\rangle$, throughout the spectrum for different ensembles.  This 
	turns out not only system-dependent but also sensitive to 
	the numerical values of the variances of the matrix elements distribution 
	\textit{i.e.}  again strongly depending on ensemble constraints. 
	IPR corroborates the belief of Poisson like behavior (\textit{i.e.} 
	localized eigenfunctions) at the edges while more delocalized 
	eigenfunctions in the bulk of the spectrum. 		
	The rescaled complexity parameter $\Lambda$ as we introduced in earlier 
	section is also strongly dependent on the local spectral energy range as well as the ensemble constraints which is clear from its definition (Eq.~\ref{9}) (also see Appendix \ref{sA}).

	
	Among various spectral fluctuation measures, ratio of nearest neighbor 
	spacing has gained a lot of popularity due to its insensitivity to the 
	otherwise difficult procedure of unfolding 
	\cite{oganesyan2007localization, atas2013distribution, 
		atas2013joint, corps2020distribution}. It is defined as,
	\begin{equation}
	\label{ri}
	r_i=s_{i+1}/s_i
	\end{equation}
	where $s_i=e_{i+1}-e_i$ is the distance between two consecutive 
	eigenvalues and the distribution of $r$ can be denoted as 
	\begin{equation}
	P(r)=\sum_i \langle \delta(r-r_i)\rangle
	\end{equation}
	For the spectral statistics in the Poisson and Wigner-Dyson
	limit, $P(r)$ can be given as
	\begin{equation}
	P(r)=
	\begin{cases}
	\frac{1}{(1+r)^2} & \text{for Poisson}\\
	
	\frac{27}{8} \frac{(r+r^2)}{(1+r+r^2)^{(5/2)}}  & \text{for GOE}.
	\end{cases}
	\end{equation}\\
	
\begin{figure}[ht!]
		\centering
		\includegraphics[width=1.0\linewidth]{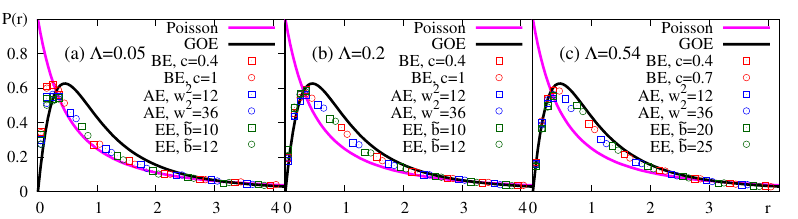}
		\caption{$P(r)$	compared for all the three ensembles for two different variances considered in each case together. Almost overlap of the curves verifies universality based on $\Lambda$.}   
		\label{fig-sp}
	\end{figure}

	We plot $P(r)$ at \new{three arbitrarily chosen} values of rescaled complexity 
	parameter $\Lambda$ for all the three type of ensembles with two 
	different choices of variances for each case in figure \ref{fig-sp}. 
	The energy ranges are different for different cases. The plot shows an 
	universal behavior of nearest neighbor spacing distribution for 
	different ensemble constraints as well as for different type of 
	ensemble with same matrix constraints. Another important result here 
	to notice is that \new{with the increasing} numerical value of $\Lambda$, the 
	statistics shifts to more chaotic side while a comparatively smaller 
	value of $\Lambda$ showing \new{statistics near Poissonian behavior}. 
	This motivated us to study the spectral behavior across the range of 
	rescaled complexity 
	parameter $\Lambda$.

	We can see from Eq.~\ref{ri} that $ r_i $ is an unbounded variable. 
	Therefore, we use a 
related measure $ \tilde{r}_i $ defined as,
\begin{equation}
\tilde{r}_i={\rm min}\Big(r_i, \frac{1}{r_i}\Big).
\end{equation}
The limiting values of ensemble averaged $ \tilde{r}_i $, denoted 
henceforth by $\langle\tilde{r}_i\rangle$ for integrable and 
chaotic systems are as follows \cite{atas2013distribution}:
\begin{equation}
\langle\tilde{r}_i\rangle=
\begin{cases}
2\;{\rm ln}(2) -1 \approx 0.3863 & \text{for Poisson}\\
4-2\sqrt(3) \approx 0.5359 & \text{for GOE}.
\end{cases}
\end{equation}

\begin{figure}[ht!]
			\centering
		\includegraphics[width=0.75\linewidth]{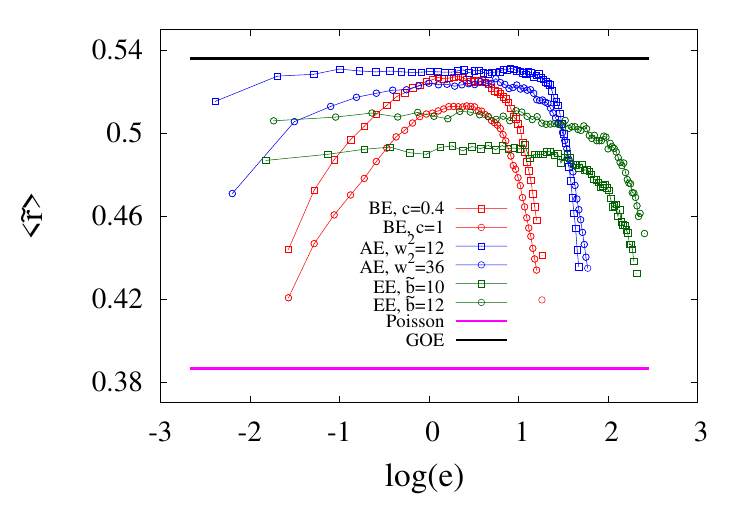}
		\caption{$\langle\tilde{r}_i\rangle$ with respect to $\log(e_i)$ (for $e_i>0$) for 
			(a) BE, (b) AE and (c) EE for two different variances considered in 
			each case. The behavior is strongly system-dependent and sensitive to the value of the variances as well.}
		\label{fig-re}
	\end{figure}

	We study energy resolved $\langle\tilde{r}_i\rangle$ to bring a fore the 
	energy dependence  of nature of spectrum.
	Figure~\ref{fig-re} depicts the variation of 
	$\langle\tilde{r}_i\rangle$ as a function of $\log(e_i)$ where $e_i$ varies 
	from center to the edge of the spectrum since it is sufficient to 
	consider  either half of the spectrum due to particle-hole symmetry. 
	For different ensembles, 
	the variation of $\langle\tilde{r}_i\rangle$ is different and also with 
	the change of the value of the variance of the matrix elements 
	distribution, dependence of $\langle\tilde{r}_i\rangle$ on the energy 
	changes. Commensurate with the general belief, edge is behaving more like 
	an integrable spectrum while bulk is similar to chaotic despite choosing a 
	purely chaotic ensemble. But what is more interesting is the dependence on 
	the variances \textit{i.e.} ensemble constraints. 
	The effect of ensemble constraint on the spectrum is clear from 
	density of states plots along with ratio of nearest neighbor spacing 
	plot and this effect on the eigenstates is prominent from IPR plots.
	
	The localized and delocalized behavior at different energy range for 
	both the ratio of spacing \textit{i.e.} eigenvalues property and IPR 
	which is an eigenfunction property 
		 brings us naturally to the question whether 
		 a transition from Poisson to Wigner like behavior 
		 can be studied in a more unified manner combining the effects of spectrum as well as the ensemble constraints?
		 
		\begin{figure}[H]
		\centering
		\includegraphics[width=0.75\linewidth]{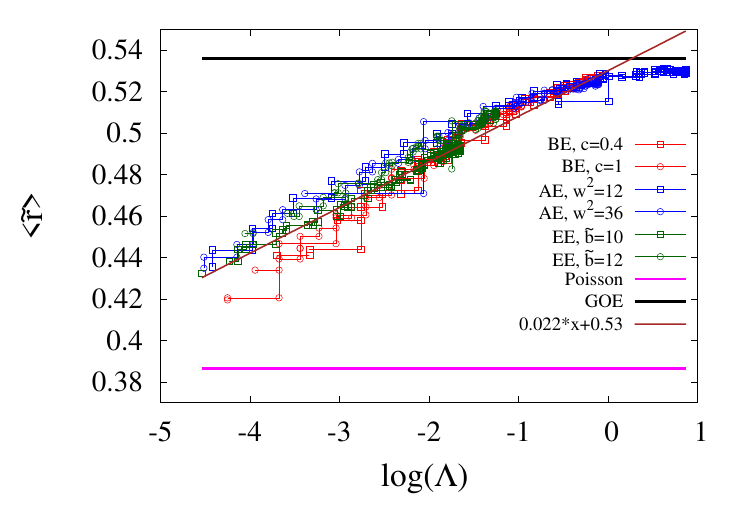}
		\caption{Dependence of $\langle\tilde{r}_i\rangle$ on $\Lambda(e_i)$ 	compared for all the three ensembles for two different variances 
			considered in each case together. $\langle\tilde{r}_i\rangle$ is 
			almost system independent verifying universality based on $\Lambda$.}
		\label{fig6}
	\end{figure}	 
	To answer this question we plotted the \new{ensemble averaged} ratio of nearest 
	neighbor spacing 
	against the rescaled complexity parameter. It shows an universal transition 
	from Poisson to Wigner\new{-Dyson} like behavior as seen in figure~\ref{fig6}.
	Universality here is meant for not only the different ensemble 
	constraints, but also the different type of ensembles keeping the matrix 
	constraint same. This transition is clearly logarithmic in re-scaled 
	complexity parameter $ \Lambda $ and $\langle\tilde{r}_i\rangle$ saturates 
	after a certain value of $\Lambda$. Previous studies for systems with and 
	without chiral symmetry \cite{mondal2020spectral, shukla2005level} show 
	this universality of the complexity parameter. Our results for different 
	ensembles with particle-hole symmetry enforces further the universal 
	behavior of $\Lambda$. However, lack of the detailed knowledge of exact 
	dependence of $\Lambda$ on the localization of eigenstates (depicted 
	through $\zeta$ in the definition of $\Lambda$ (Eq.~\ref{33}) is most 
	probably the reason for the slight deviation of 
	$\langle\tilde{r}_i\rangle$ from each other in figure~\ref{fig6}).

	\section{Interpolating ensemble from particle-hole to chiral 
		symmetry}
	\label{s7}

	Motivated by the universality found in previous section and logarithmic 
	dependence of $ \langle \tilde{r} \rangle$ on $ \Lambda $, we introduce an 
	interpolating ensemble which connects a system 
	with particle-hole symmetry to one with chiral symmetry. Let us recall that 
	in case of chiral symmetric Hamiltonian the diagonal blocks in Eq.~\ref{5} are 
	zero matrices of appropriate dimension.
	
	Previously, transition from Poisson to GOE 
	\cite{bohigas1993manifestations, french1988statistical, wigner1967random, 
		cerruti2003uniform, guhr1996transition}, 
	appearance of partial transport barriers in transport 
	\cite{bohigas1993manifestations, michler2012universal}, perturbation of 
	integrable dynamical system \cite{bohigas1993manifestations, 
		berry1984semiclassical}, in the coupled chaotic 
	systems\cite{srivastava2016universal}, effect of disorder strength 
	in 
	Anderson model\cite{evers2008anderson}; GOE to GUE transition 
	\cite{pandey1983gaussian, bohigas1995chaotic} all have been studied in the 
	framework of dynamical symmetry breaking in various interpolating 
	ensembles. In contrast to this, here we are 
	interested in Poisson to GOE transition of spectral fluctuations within a 
	single spectrum for each instance of interpolating ensembles.

	The interpolating Hamiltonian is defined as
	\begin{equation}
	H_{inter}=
	\begin{pmatrix}
	\epsilon\mathcal{H} & \Delta\\
	\Delta & -\epsilon\mathcal{H}
	\end{pmatrix}
	\label{epsi}
	\end{equation}
	where $ \epsilon \in [0,1] $ is the interpolating parameter defining the 
	chiral symmetric and particle-hole symmetric systems at the end points. As 
	the matrix elements are chosen as Gaussian distributed with zero mean and unit variance, 
	multiplication of $ \epsilon $ scales the variance of diagonal block 
	elements by $ \epsilon^2 $. We take $\epsilon=1, 10^{-1}, 10^{-2}, 0$ as 
	the representative values. $\epsilon =0$ leads to a system with 
	\textit{symmetric} chiral symmetry:
	\begin{equation}
	H_c=
	\begin{pmatrix}
	0 & \Delta\\
	\Delta & 0
	\end{pmatrix}
	\label{57}
	\end{equation}
	It is symmetric because the off-diagonal blocks $\Delta$ are individually 
	symmetric in nature and therefore, reducing the number of independent 
	matrix elements to $N(N+1)/2$ for a $2N\times 2N$ Hamiltonian matrix $H_c$. Note that, 
	in general, the $\Delta$ blocks are not necessarily symmetric or even 
	square for chiral systems.

	Now to calculate the complexity parameter $Y$ as well as $\Lambda$ we 
	again consider the three different types (BE, AE and EE) of interpolating 
	ensembles by taking the variances of the matrix elements distribution 
	differently.
	\subsection{Brownian ensemble}
	Considering the variance of the diagonal elements of $\mathcal{H}$ to be
	\begin{align}
	h_{kk}=\frac{\epsilon^2}{2\gamma}
	\label{beh}
	\end{align}
	and keeping the other parameters as they were in Eq.~\ref{19}, 
	\ref{20} and \ref{21} (for $H$ in Eq.~\ref{5}) and following the same equation of complexity parameter provided in section \ref{s5a}, we 
	achieve 
	\begin{align}
	Y-Y_0\simeq \frac{1}{4\gamma \mu} (1+\epsilon^2).
	\label{bey}
	\end{align}
	The correlation volume $\zeta$ in the definition of the rescaled complexity parameter $\Lambda$ (Eq.~\ref{33}) was considered to be same as the size ($2N$) of the Hamiltonian matrix in case of Brownian ensemble in \ref{s5a}. But $\epsilon\leq 0.1$ makes $\zeta$ to be dependent on $\epsilon$ as $\zeta = 2N(1-5^{\alpha}\epsilon)$ where $\epsilon=10^{-\alpha}$ with $\alpha$ to be non-zero positive integer. Therefore, $\Lambda$ becomes 
	\begin{equation}
	\Lambda_B(e,Y)=(Y-Y_0)((1-5^{\alpha}\epsilon)R_1(e))^2
	\label{be-epsi}
	\end{equation}
	with $(Y-Y_0)$ provided by Eq.~\ref{bey}.
	\subsection{Anderson ensemble}
	If the variance of the elements of $\mathcal{H}$ are considered to be
	\begin{align}
	h_{kk}=\frac{\epsilon^2 w^2}{12},\qquad
	h_{kl}=\epsilon^2\frac{w_s^2}{12}\quad (k,l \;{\rm for}\;{\rm nn}\;{\rm 
		sites})
	\label{aeh}
	\end{align}
	and the other multi-parameters are defined as they were in Eq.~\ref{h} and 
	\ref{b} (for $H$ in Eq.~\ref{5}), following the equation of 
		complexity parameter provided in section \ref{s5b}, one can get
	\begin{align}
	Y-Y_0= -\frac{2}{N+1}\;{\rm ln} \Big\{\Big |1-\frac{w^2}{48}\Big|\Big 
	|1-\frac{\epsilon^2 w^2}{48}\Big|\Big\}
	-\frac{4}{N+1}\;{\rm 
		ln}\Big\{\frac{1}{2}\Big(1-\frac{\epsilon^2}{2}\Big)\Big\}
	\label{aey}
	\end{align}
	The correlation volume $\zeta$ here consists of IPR as discussed in section \ref{s5b} which takes care the effect of $\epsilon$ on the localization of wavefunctions. The rescaled complexity parameter $\Lambda$, therefore, will be same as defined in Eq.~\ref{44a} with $(Y-Y_0)$ given by Eq.~\ref{aey}.
	
	\subsection{Ensemble with exponential decay}
	The variance of the elements of $\mathcal{H}$ for EE is given by 
	\begin{align}
	h_{kl}=\frac{\epsilon^2}{{\rm exp}\big(\frac{|k-l|}{\tilde{b}}\big)^2}
	\label{eeh}
	\end{align}
	for $k,l=1\to N$ and the variances of the other elements distribution are 
	given by Eq.~\ref{45} (for $H$ in Eq.~\ref{5}) with the mean of all 
	elements distribution to be zero. To calculate the complexity parameter, 
		if we follow the equation provided in section \ref{s5c}, we get
	\begin{align}
	Y-Y_0=-\frac{4}{N(N+1)}\sum_{r=1}^{N-1}(N-r)\;{\rm ln} 
	\Big\{\Big|1-\frac{1}{2\;{\rm 
			exp}(\frac{r}{\tilde{b}})^2}\Big|
	\Big|1-\frac{\epsilon^2}{2\;{\rm exp}(\frac{r}{\tilde{b}})^2}\Big|\Big\}.
	\label{eey}
	\end{align}

	\begin{figure}[H]
		\centering
		\includegraphics[width=0.52\linewidth]{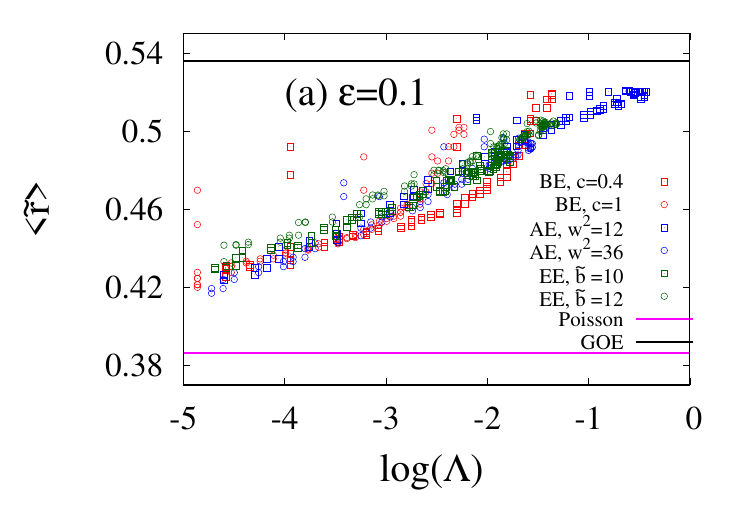}
		\hspace{-0.8cm}
		\includegraphics[width=0.52\linewidth]{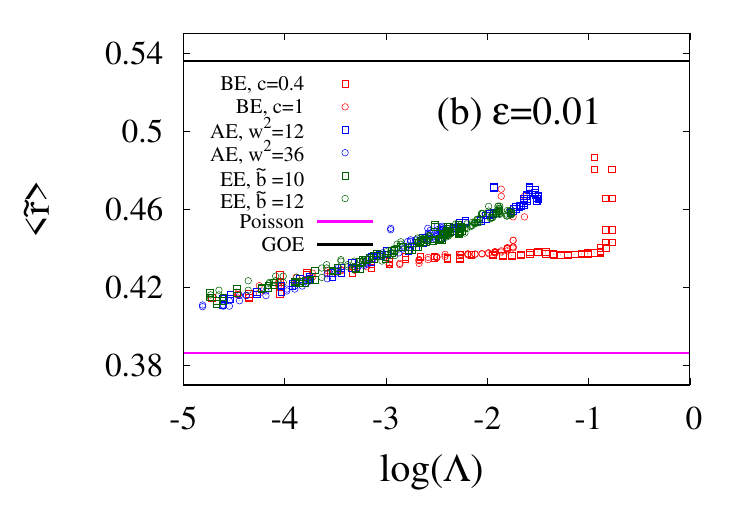}
		\includegraphics[width=0.52\linewidth]{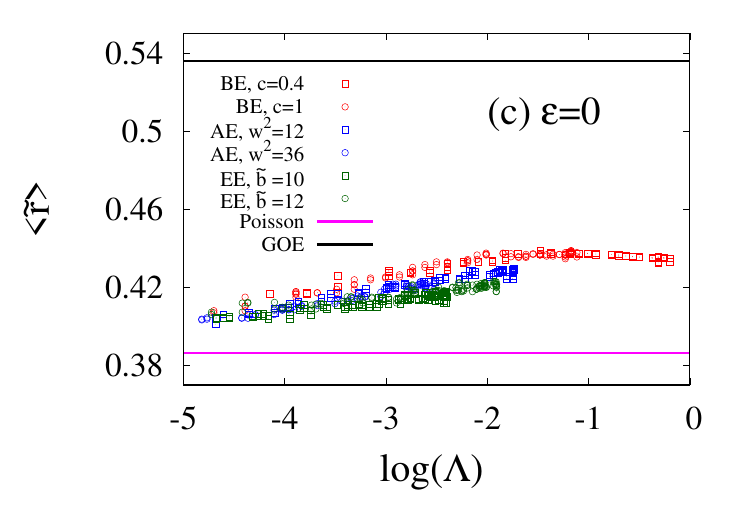}
		\caption{$\langle\tilde{r}_i\rangle$ vs $\log(\Lambda)$ plot for (a) 
			$\epsilon=10^{-1}$, (b)  $\epsilon=10^{-2}$ and (c) $\epsilon=0$. 
			The 
			$\langle\tilde{r}_i\rangle$ for different variances of same type 
			of 
			ensembles as well as for different ensembles coincide with each 
			other during the transition from Poisson 
			$\to$ GOE before they saturate. As the 
			value of the parameter $\epsilon$ decreases (\textit{i.e.} goes towards chiral ensemble from particle-hole one), $\langle\tilde{r}_i\rangle$ saturates faster for all the cases.}   
		\label{fig7}
	\end{figure}
		In this case, $\zeta$ is again a function of IPR provided in section \ref{s5c} and hence, not explicitly dependent on $\epsilon$ which keeps the definition of the $\Lambda$ as it was defined in Eq.~\ref{49} with $(Y-Y_0)$ to be as Eq.~\ref{eey}.
		
	The variation of $\langle\tilde{r}_i\rangle$  with $\Lambda$ for different $ \epsilon $  as 	well as for different ensembles is shown in figure~\ref{fig7}. For $ \epsilon=0 $, the system goes away from the integrability as a function of $ \log(\Lambda) $ but saturates before it reaches to chaotic limit. This incomplete transition can be attributed to lack of sufficient independent elements in the Hamiltonian due to added symmetric nature of off-diagonal blocks. During transition, the independence on ensemble constraint once again displays the universal nature of the 
	transition within the spectrum. With increasing value of $ \epsilon $, 
	complete transition happens and the universality in the independence of 
	ensemble constraint as well as the variation of $ 
	\langle \tilde{r} \rangle \propto \log(\Lambda)$ is maintained.

	\section{Application to 2D SSH model}
	\label{s9}
\begin{figure}[ht!]
		\centering
		\includegraphics[width=0.4\linewidth]{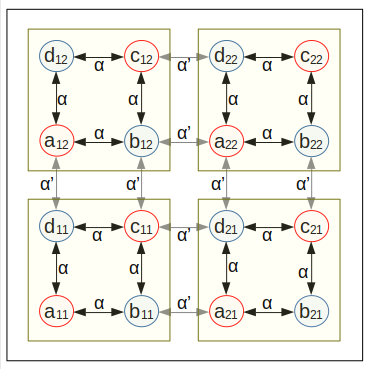}
		\hspace{0.5cm}
		\includegraphics[width=0.4\linewidth]{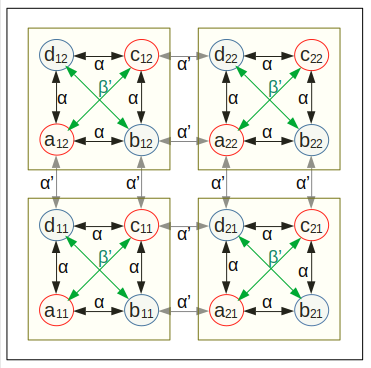}
		\caption{A 2D SSH model for lattice size $L=4$ is shown here. 
		The \textit{red} and the \textit{blue} lattice points 
		represent their associated sublattices to be different from each 
		other. The \textit{yellow} squares represent each unit cell 
		consisting four lattice points; two from each sublattice. The 
		\textit{black} arrows represent interaction between two lattice 
		points belonging to different sublattice inside ($\alpha$) the 
		unit cells whereas the \textit{grey} ones represent that outside 
		($\alpha^{\prime}$) the unit cells. Left figure is for 2D SSH 
		model with chiral symmetry and the right one is that with 
		particle-hole symmetry. The latter one also consists diagonal 
		interactions within the same unit cell represented by 
		\textit{green} arrows.}
		\label{fig11}
	\end{figure}

	In this section, we explore this universality for a physical 
	example of 2D SSH like model \new{once with chiral symmetry and again with particle-hole symmetry}.
	This two dimensional generalization of 1D SSH model is a 
	representation of bipartite lattice structure where the interaction strength inside and outside of the unit cell are considered to be different from each other. 
	Figure~\ref{fig11} is a diagrammatic representation of the details of the 
	model where  two sublattices A and B having $a11,a21,a31,\dots\new{,c11,c21,c31,\dots}$ and $b11,b21,b31,\dots\new{,d11,d21,d31,\dots}$ lattice points respectively are shown along with the unit cell which consists of four lattice points \new{(one of each $a,b,c$ and $d$). The pair of lattice points belonging to the same sublattice \new{((a and c) or (b and d))} sit at the ends of the diagonal of a square unit cell}. 
		
\new{	
	\subsection{2D SSH model with chiral symmetry}
Bipartite nature of the SSH model makes the nearest neighbors (both inside 
and outside of the unit cell) of every lattice point to belong to the 
other sublattice.  Only nearest neighbor interactions are being 
considered} in this 
representation of 2D SSH model \new{ with chiral symmetry [Left figure of 
Fig.~\ref{fig11}]. Here the interaction between the lattice points 
belonging to same sublattice are forbidden}. \new{The interaction strength 
inside and outside of the unit cell are taken different from each other 
$\alpha$ and $\alpha^{\prime}$ respectively.
	The interaction Hamiltonian of this model looks like 
	\begin{eqnarray}
	H_S^{ch}=\sum_{x,y=1}^{L/2}&\Bigl[&\alpha_{xy}^{ab}\;a^\dagger_{xy}b_{xy}+\alpha_{xy}^{cd}\;c^\dagger_{xy}d_{xy}+\alpha_{xy}^{da}\;d^\dagger_{xy}a_{xy}+\alpha_{xy}^{bc}\;b^\dagger_{xy}c_{xy}\Bigr.\nonumber\\
	&&\Bigl.\alpha^{\prime ba}_{xy}\;b^\dagger_{xy}a_{x+1,y}+\alpha^{\prime cd}_{xy}\;c^\dagger_{xy}d_{x+1,y}+\alpha^{\prime da}_{xy}\;d^\dagger_{xy}a_{x+1,y}+\alpha^{\prime cb}_{xy}\;c^\dagger_{xy}b_{x+1,y}\Bigr]\nonumber\\
	\label{hssh}
	\end{eqnarray} 
	with $L$ being the lattice size.
	The Hamiltonian $H_S^{ch}$ can be expressed as $H_S^{ch}=\psi^\dagger 
	H_{SSH}^{ch}\psi$ with }
	\begin{equation}
	H_{SSH}^{ch}=
	\begin{pmatrix}
	0 & \Delta\\
	\Delta^T & 0
	\end{pmatrix}
	\label{ssh}
	\end{equation}
	in site basis \new{$\psi^\dagger=(a^\dagger_{11},a^\dagger_{21},\dots,c^\dagger_{11},c^\dagger_{21},\dots,b^\dagger_{11},b^\dagger_{21},\dots,d^\dagger_{11},d^\dagger_{21},\dots)$}. Here, the off diagonal block matrix $\Delta$ is of size 
	$N\times N$ for 
	\new{ $H_{SSH}^{ch}$ of size $2N=L^2$}. The $H_{SSH}^{ch}$ is a 
	 representation of systems with chiral symmetry where the 
	off-diagonal blocks are not symmetric. 
The energy spectrum in this case \new{like particle-hole symmetric system} is symmetric around zero.
	
	For numerical analysis \new{of spectral statistics}, we have considered an ensemble of $500$ matrices 
	of 2D lattice of \new{size $L=32$}. We defined the interaction 
	strengths inside and outside 
	($\alpha$ and $\alpha^{\prime}$ respectively) of the unit cell as Gaussian 
	distributed random number to model the disorder. The mean and variance are 
	the relevant parameters for the present study. Here we have considered 
	three combinations of \new{the variances} $h_\alpha$ and $h_{\alpha'}$:  i) 
	$h_{\alpha}=1,h_{\alpha^{\prime}}=3/2$, ii) 
	$h_{\alpha}=3/2,h_{\alpha^{\prime}}=1$ and iii) 
	$h_{\alpha}=3,h_{\alpha^{\prime}}=1$ once with zero mean of the matrix 
	elements and at other time with non-zero mean $ b_\alpha = 0.2, 
	b_{\alpha'} 
	= 0.1 $.
	The density of states is symmetric around zero 
	energy as expected due to chiral symmetry. The level density 
	depends on the sum of the variances of interaction strengths inside and 
	outside of the unit cell. The width of the spectrum increases with the 
	increasing value of $h_\alpha + h_{\alpha'}$ while IPR
	$\langle I_2\rangle$ depends on the sum as well as the modulus difference 
	between $h_{\alpha}$ and $h_{\alpha^{\prime}}$. For a fixed value of $ 
	\lvert h_\alpha - h_{\alpha'}\rvert $, with increasing value of $h_\alpha 
	+ h_{\alpha'}$, the eigenstates tend to delocalize. However, 
	for a fixed sum, localization increases with increasing $ 
	\lvert h_\alpha - h_{\alpha'}\rvert $ which is understandable as the 
	strength of the disorder become at par inside and outside the unit cell.
	
\begin{figure}[ht!]
		\centering
		\includegraphics[width=0.52\linewidth]{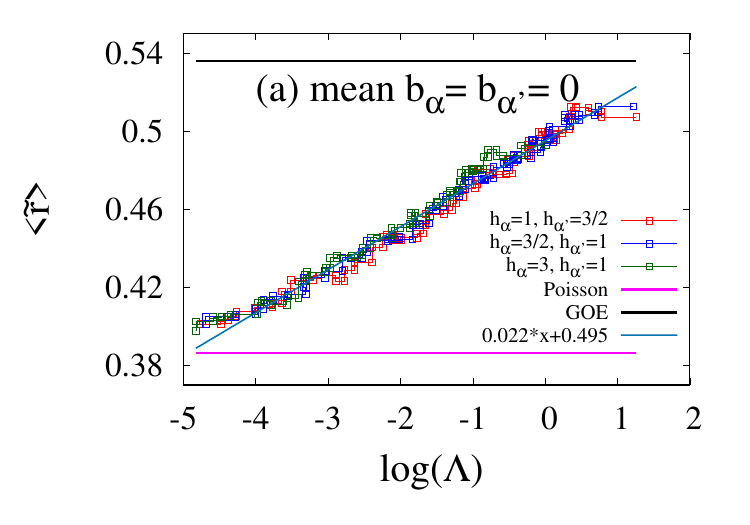}
		\hspace{-0.8cm}
		\includegraphics[width=0.52\linewidth]{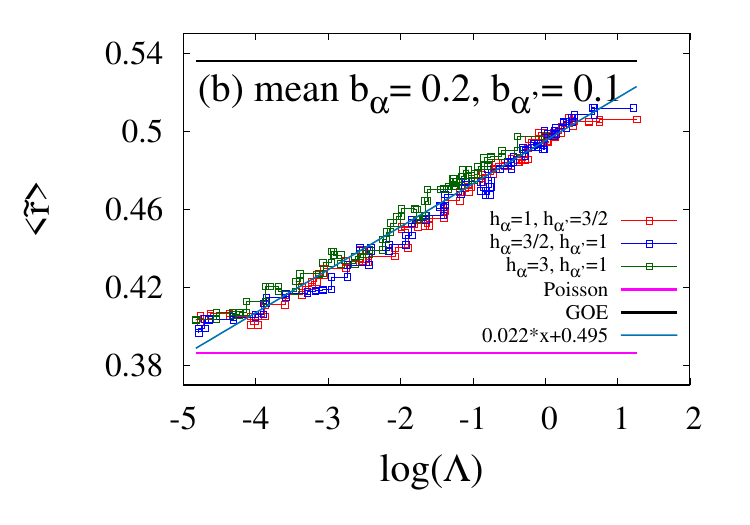}
		\caption{$\langle\tilde{r}_i\rangle$ with respect to $\log(\Lambda(e_i))$ 
			for 2D SSH \new{with chiral symmetry} for different distribution of the interaction strengths $\alpha$ and $\alpha^{\prime}$.  In (a) mean of the distribution $ b_\alpha = b_{\alpha'} = 0 $ whereas in (b) $ b_\alpha = 0.2, b_{\alpha'} = 0.1 $. For both the cases three different combinations of the variances are plotted. With the increasing \new{numerical} value of $\Lambda$, the $\langle\tilde{r}_i\rangle$ goes from \new{near} Poisson to \new{close to} GOE limit and this transition is independent of all the ensemble constraints considered like means and variances.}
		\label{fig13}
	\end{figure}

	Following the definition of the complexity parameter provided in 
	Eq.~\ref{8}, the $Y$ in this case is given by 
	\begin{align}
	Y = -\frac{1}{2M\gamma}\Big[N\; {\rm ln} \Big |1-\gamma h_{\alpha}\Big| 
	+ N \;{\rm ln} \Big |1-2\gamma h_{\alpha}\Big| + 2N \;{\rm ln} \Big |1-2\gamma h_{\alpha^{\prime}}\Big| \nonumber\\
	 + 2N \;{\rm ln}\;|b_{\alpha}|^2 +2N \;{\rm ln}\;|b_{\alpha'}|^2 \Big]+Y_0 
	\label{sshy}
	\end{align}
	where $M$ is the number of non-zero independent matrix elements, $\gamma$ 
	is an arbitrary parameter considered to be equals to $1/4$ throughout 
	while $ Y_0 $ is the complexity parameter of the initial state which 
	corresponds to the Poisson limit attained by the system. For our 
	calculation, we have chosen interaction inside the unit cell much 
	higher ($h_{\alpha}=9$) than that at outside ($h_{\alpha^{\prime}}=0.01$) 
	for this limit. 
	In total eight logarithmic terms appear as we consider two dimensional system 
	resulting in the number of the nearest neighbor elements to be four. 
	Considering $M=N(N+1)/2$ and the correlation volume 
	$\zeta$ to be equal to the participation ratio (\textit{i.e.}
	$\zeta=\frac{1}{\langle I_2\rangle}$), the rescaled complexity parameter 
	$\Lambda$ becomes same as Eq.~\ref{44a}, where $Y-Y_0$ is defined by Eq.~\ref{sshy}. The \new{ensemble averaged} ratio of nearest neighbor spacing $ \langle \tilde{r} \rangle $ is plotted 
	as function of $ \log(\Lambda) $ in figure~\ref{fig13} for this 2D SSH model \new{with chiral symmetry}. 

	It is clear from figure~\ref{fig13} that $\langle\tilde{r}_i\rangle$ not 
	only shows a transition from nearly integrable value to chaotic limit with 
	the increase of the numerical value of $\Lambda$ but also this 
	transition is independent of the distribution parameters (mean and 
	variances which are ensemble constraints), verifying universality of 
	spectral fluctuations based on $\Lambda$. The logarithmic 
	dependence of \new{averaged} nearest neighbor spacing ratio on complexity parameter 
	agrees well with our result described in section \ref{s6}. This 
	correspondence in first glance may look surprising as true 
	symmetry of 2D SSH model considered here is chirality however the 
	resolution of this is 
	in the total number of ensemble constraints which in this case is same as 
	in the system with particle-hole symmetry.  The mean value 
	of the matrix elements distribution 
	\new{does not change the logarithmic} dependence of $ \langle\tilde{r}_i\rangle $ on the numerical values of $ \Lambda $, 
	moreover the rate with which this transition happens is independent of mean.
	
	\new{\subsection{2D SSH model with CI class of particle-hole symmetry}
	In the particle-hole symmetric version of 2D SSH model, along with 
	already existing interactions, we now consider onsite disorder of 
	strength $\beta$ for all lattice points and non-zero interaction of 
	strength $\beta^\prime$ between the lattice points belonging to same 
	sublattice only inside the unit cells (a-c and b-d) [Right figure of 
	Fig.~\ref{fig11}]. The interaction Hamiltonian will now become 
	\begin{eqnarray}
	H_S^{ph}=H_S^{ch}+\sum_{x,y=1}^{L/2}&\Bigl[&\beta_{xy}^{aa}\;a^\dagger_{xy}a_{xy}+\beta_{xy}^{bb}\;b^\dagger_{xy}b_{xy}+\beta_{xy}^{cc}\;c^\dagger_{xy}c_{xy}+\beta_{xy}^{dd}\;d^\dagger_{xy}d_{xy}\Bigr.\nonumber\\
	&& +\beta_{xy}^{\prime ac}\;a^\dagger_{xy}c_{xy}+\beta_{xy}^{\prime bd}\;b^\dagger_{xy}d_{xy}\Bigr]
	\label{hsshp}
	\end{eqnarray} 
	with $H_S^{ch}$ taken from Eq.~\ref{hssh}. The Hamiltonian $H_S^{ph}$ 
	can be expressed as $H_S^{ph}=\psi^\dagger H_{SSH}^{ph}\psi$ with 
	$\psi^\dagger=(a^\dagger_{11},a^\dagger_{21},\dots,c^\dagger_{11},c^\dagger_{21},\dots,b^\dagger_{11},b^\dagger_{21},\dots,d^\dagger_{11},d^\dagger_{21},\dots)$
	 where the Hamiltonian matrix $H_{SSH}^{ph}$ will have non-zero 
	diagonal blocks. It will be particle-hole symmetric of CI class, if 
	$H_S^{ph}$ has additional terms: $\sum_{x,y=1}^{L/2}[\alpha^{\prime 
	ba}_{xy}\;b^\dagger_{x+1,y}a_{xy}+\alpha^{\prime 
	cd}_{xy}\;c^\dagger_{x+1,y}d_{xy}]$, along with the conditions as 
	follows
\begin{eqnarray}
\alpha_{xy}^{da}=\alpha_{xy}^{bc},\quad
\alpha_{xy}^{\prime da}=\alpha_{xy}^{\prime cb},\quad
\beta_{xy}^{aa}=-\beta_{xy}^{bb},\quad
\beta_{xy}^{cc}=-\beta_{xy}^{dd}\quad
{\rm and}\quad \beta_{xy}^{\prime ac}=\beta_{xy}^{\prime bd}.
\end{eqnarray}
$H_{SSH}^{ph}$ will now exactly look like $H$ in Eq.~\ref{5} with same number of independent ensemble constraints.

For numerical study, all the interaction strengths ($\beta$ and $ \beta^{\prime}$ along with $\alpha$ and $\alpha^{\prime}$) are defined to be Gaussian distributed random numbers. Four combinations of the variances $h_\alpha, h_{\alpha'}, h_\beta$ and $h_{\beta'}$ are considered: i) $h_\alpha=h_{\alpha'}=h_\beta=h_{\beta'}=1$, ii) $h_\alpha=h_{\alpha'}=1, h_\beta=h_{\beta'}=3/2$, iii)  $h_\alpha=h_{\alpha'}=3/2, h_\beta=h_{\beta'}=1$ and iv)  $h_\alpha=h_\beta=1, h_{\alpha'}=h_{\beta'}=3/2$ for both zero and non-zero mean $b_\alpha=b_\beta=0.1, b_{\alpha'}=b_{\beta'}=0.2$.

The complexity parameter in this case, following the definition in Eq.~\ref{8}, will now become $Y=Y_{od}+Y_d+Y_0$ with
	\begin{eqnarray}
	Y_{od}&\equiv -\frac{1}{2M\gamma}\Big[N\; {\rm ln} \Big |1-\gamma h_{\alpha}\Big| 
	+ \frac{N}{2}\;{\rm ln} \Big |1-2\gamma h_{\alpha}\Big| &+ \frac{3N}{2} \;{\rm ln} \Big |1-2\gamma h_{\alpha^{\prime}}\Big| \nonumber\\
	&& + \frac{3N}{2} \;{\rm ln}\;|b_{\alpha}|^2 + \frac{3N}{2} \;{\rm ln}\;|b_{\alpha'}|^2 \Big]
	\label{sshyo}\\
	Y_{d}&\equiv -\frac{1}{2M\gamma}\Big[N\; {\rm ln} \Big |1-\gamma h_{\beta}\Big| 
	+ \frac{N}{2}\;{\rm ln} \Big |1-2\gamma h_{\beta'}\Big| &+ N \;{\rm ln}\;|b_{\beta}|^2 + \frac{N}{2} \;{\rm ln}\;|b_{\alpha'}|^2 \Big]
	\label{sshyd}
	\end{eqnarray}
	with $M$ and $\gamma$ being again the number of non-zero independent matrix elements considered as $M=N(N+1)/2$  and an arbitrary parameter to be equal to $1/4$ respectively. $ Y_0 $ is the complexity parameter at the initial state. We have chosen the onsite disorder much higher ($h_{\beta}=9$) than the interaction between any two lattice points ($h_{\alpha}=h_{\alpha^{\prime}}=h_{\beta^{\prime}}=0.01$) to achieve the Poisson limit by the system. Again, considering the correlation volume $\zeta$ to be equal to the participation ratio, the rescaled complexity parameter $\Lambda$ becomes same as Eq.~\ref{44a}, where $Y-Y_0=Y_{od}+Y_{d}$ with $Y_{od}$ and $Y_{d}$ being defined by Eq.~\ref{sshyo} and Eq.~\ref{sshyd} respectively. Figure~\ref{fig14} shows the ensemble averaged ratio of nearest neighbor spacing $ \langle \tilde{r} \rangle $ with respect to $ \log(\Lambda) $ for this 2D SSH model with particle-hole symmetry of CI kind. Here an ensemble of $500$ matrices of 2D lattice of size $L=32$ is considered.
\begin{figure}[ht!]
	\centering
	\includegraphics[width=0.52\linewidth]{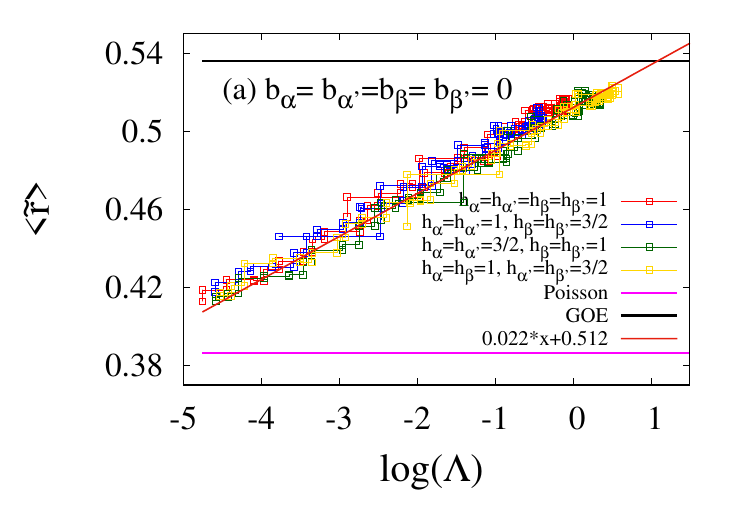}
	\hspace{-0.8cm}
	\includegraphics[width=0.52\linewidth]{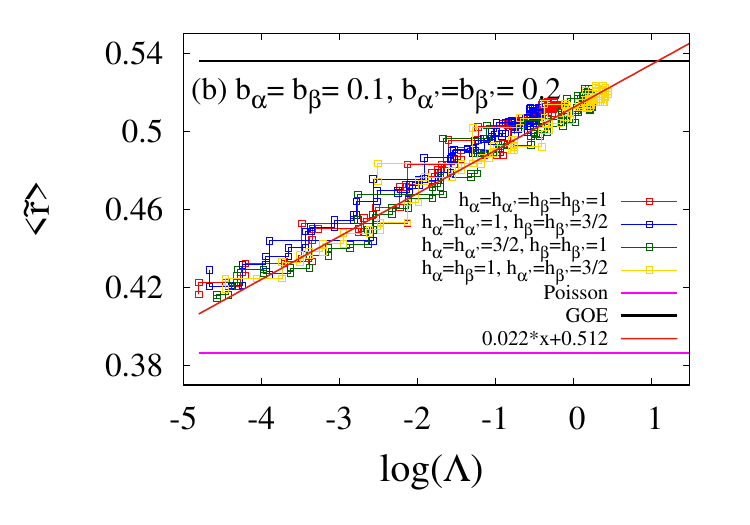}
	\caption{$\langle\tilde{r}_i\rangle$ as a function of $\log(\Lambda(e_i))$ 
		for 2D SSH with particle-hole symmetry for different distribution of the interaction strengths $\alpha,\alpha^{\prime},\beta$ and $\beta'$.  In (a) mean of the distribution $ b_\alpha = b_{\alpha'} =b_\beta = b_{\beta'} = 0 $ whereas in (b) $b_\alpha=b_\beta=0.1, b_{\alpha'}=b_{\beta'}=0.2$. For both the cases four different combinations of the variances are plotted. With the increasing numerical value of $\Lambda$, the $\langle\tilde{r}_i\rangle$ goes from near Poisson to near GOE limit and this transition is independent of all the ensemble constraints considered like means and variances.}
	\label{fig14}
\end{figure}

	Figure~\ref{fig14} again confirms a transition of $\langle\tilde{r}_i\rangle$ from nearly integrable value to chaotic limit with the increase of the numerical value of $\Lambda$ and its independence of the distribution parameters like mean and variances. This result again verifies the universality of spectral fluctuations based on $\Lambda$. Again, $\langle\tilde{r}_i\rangle \propto \log(\Lambda)$ . For both the 2D SSH models one with chiral symmetry and the other one with particle-hole symmetry, the rate of transition is same. Even the mean value does not change the logarithmic dependence of $ \langle\tilde{r}_i\rangle $ on the numerical values of $ \Lambda $, as well as the rate of the transition.
	 }
	\section{Conclusions}
	\label{s10} 
	
	In this paper, we have shown that the spectral statistics  
	characterized by \new{ensemble averaged} ratio of nearest neighbor spacing shows an universal transition 
	from integrable to chaotic behavior within single spectrum, when studied 
	as
	a function of complexity parameter for the system having particle-hole 
	symmetry.
	The universality here means that transition
	does not depend on ensemble constraints whenever the matrix constraints are
	kept fixed. Moreover, the $\langle \tilde{r}\rangle$ goes as 
	$\log(\Lambda)$. 
	The complexity parameter itself is a function of different parameters 
	present 
	in the joint probability distribution function of the system. Introduced 
	via 
	diffusion 
	equation  for JPDF in the parameter space, this quantity in addition 
	requires the local information regarding spectral density as well as localization property of 
	eigenfunctions. The different ensembles like Brownian, Anderson and 
	exponential 
	show the same behavior and completely fall on top of each other as seen 
	in 
	figure~\ref{fig6}.
	
	We further have introduced and studied an interpolating ensemble having a 
	single
	parameter $\epsilon \in [0,1]$ which gives chiral and particle-hole 
	symmetric
	Hamiltonian at its end-points. We further show that for a fixed value of 
	$\epsilon$, the 
	transition from integrable to Wigner-Dyson value follows the universality 
	and
	$\langle \tilde{r}\rangle \propto \log(\Lambda)$. Motivated by this, we 
	studied \new{two different} 2D SSH like models \new{one with chiral 
	symmetry and the other one with particle-hole symmetry} where bond 
	disorder and \new{chemical potential} are modeled by 
	Gaussian random numbers. The 
	spectral transition within the spectrum again follows the universality \new{for both the models with different combinations of mean and variances of the matrix elements distribution.} 
	
	In our analysis we have considered the localization volume $\zeta$ 
	intuitively for various ensembles guided also by previous works 
	\cite{fyodorov1994statistical, shukla2005level, mondal2020spectral}. A systematic study of this 
	as a function of energy would improve this analysis on the quantitative 
	level and can be studied in future.

	\section*{Acknowledgement} 
	One of the authors, T. Mondal, gratefully acknowledges Professor Pragya Shukla for introducing her to the complexity parameter and educating about its universal behavior for different symmetry classes of random matrix ensemble.
	
	\section*{Declarations}
\subsection*{Authors' contributions} All the authors prepared the manuscript.
\subsection*{Data availability statement} Data sharing not applicable to this article as no datasets were generated or analyzed during the current study.

\begin{appendices}
	\section{Nearest neighbor spacing distribution}
	\label{nnspacing}
	A particle-hole symmetric matrix have eigenvalues always symmetric 
	around zero. Therefore, if we sort them
	, the spacing between $e_i$ and $e_{i+1}$ will be equal to the spacing between $-e_i$ and $-e_{i+1}$, \textit{i.e.}  $|e_{i+1}-e_i|=s$.
	\begin{center}
		\includegraphics[width=0.5\linewidth]{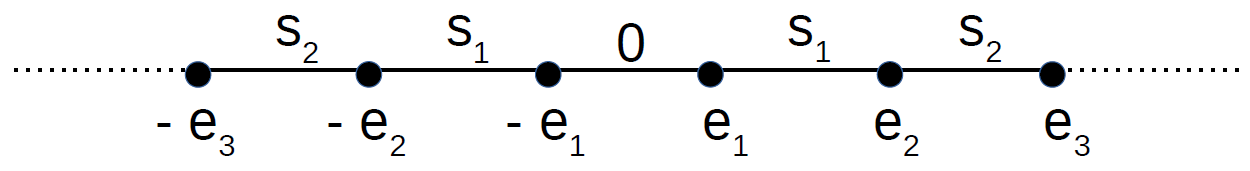}
	\end{center}
	Hence, it is sufficient to consider either side of the spectrum as long as nearest neighbor fluctuation is the spectral property of interest.
	 	
	Considering Eq.~\ref{5} to be a $4\times 4$ Hamiltonian, two of its 
	four eigenvalues will be positive due to particle-hole symmetry. 
	Therefore, without any loss of generality, restricting ourselves to 
	only positive sector of eigenvalues, we can calculate the nearest 
	neighbor spacing distribution $P(s)$.
	
	 Let us assume, the two positive eigenvalues are $e_1$ and $e_2$ and the spacing between them is $s$. Therefore, 
	 
	\begin{eqnarray}
	P(s)&=& \int_{0}^{\infty}\int_{0}^{\infty}P(e_1,e_2)\;\delta(s-|e_1-e_2|)\;d e_1de_2 \nonumber\\
	&=& \int_{0}^{\infty}de_1\int_{0}^{e_1}P_{e_1>e_2}\;\delta(s-|e_1-e_2|)\;de_2\nonumber\\&& +\int_{0}^{\infty}de_2\int_{0}^{e_2}P_{e_2>e_1}\;\delta(s-|e_2-e_1|)\;de_1 \nonumber\\
	\end{eqnarray}

Using Eq.~\ref{p}, one can write $P(e_1,e_2)$. Now,
\begin{eqnarray}
P(s)&\propto& \int_{0}^{\infty}de_1\int_{0}^{e_1}(e_1^2-e_2^2)\;e_1e_2\; \exp(-\frac{e_1^2+e_2^2}{h^2})\;\delta(s-|e_1-e_2|)\;de_2\nonumber\\ &&+\int_{0}^{\infty}de_2\int_{0}^{e_2}(e_2^2-e_1^2)\;e_1e_2\; \exp(-\frac{e_1^2+e_2^2}{h^2})\;\delta(s-|e_2-e_1|)\;de_1 \nonumber\\
&=& 2 \int_{0}^{\infty}de_1(e_1^2-(e_1-s)^2)\;e_1(e_1-s)\; \exp(-\frac{e_1^2+(e_1-s)^2}{h^2}) \nonumber\\
&=&\frac{h^4}{2}\;s\;e^{-s^2/h^2} 
\end{eqnarray}

Therefore,\begin{equation}
 P(s)\propto s\;e^{-s^2/h^2}.
\end{equation}
This spacing distribution is same as that of GOE. Now for $2000\times 2000$ Hamiltonian matrix we numerically verified this analytical result.
\begin{figure}[ht!]
	\centering
	\includegraphics[width=0.52\textwidth]{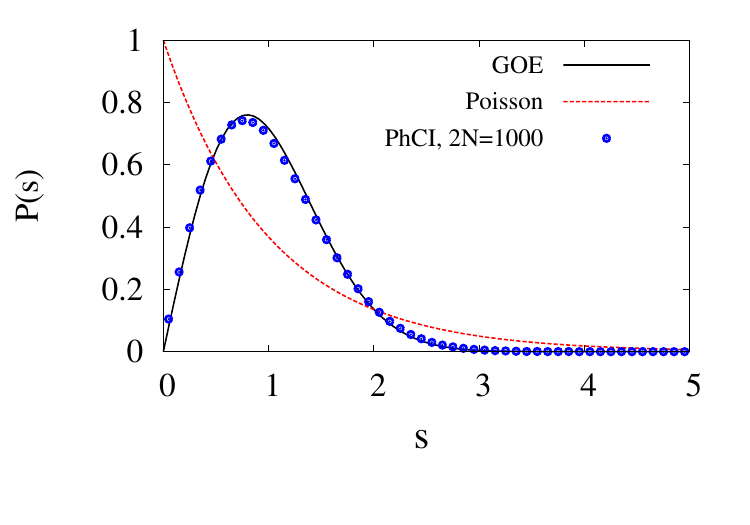}
	\caption{Distribution of nearest neighbor spacing for an ensemble of $5000$ Ph-CI matrices of size $2000\times 2000$ }
	\label{figPs}
\end{figure}

Since our spectral 
property of interest is nearest neighbor spacing ratio distribution, we 
numerically present here $P(r)$ and $P(\tilde{r})$ for particle-hole CI 
class of random matrices with no additional constraints (Eq.~\ref{5}),
\begin{figure}[ht!]
	\centering
	\includegraphics[width=0.52\textwidth]{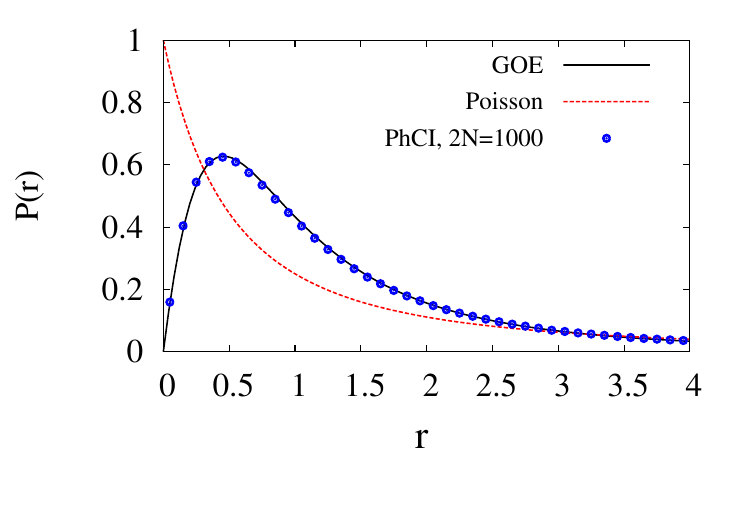}
	\hspace{-0.8cm}
	\includegraphics[width=0.52\textwidth]{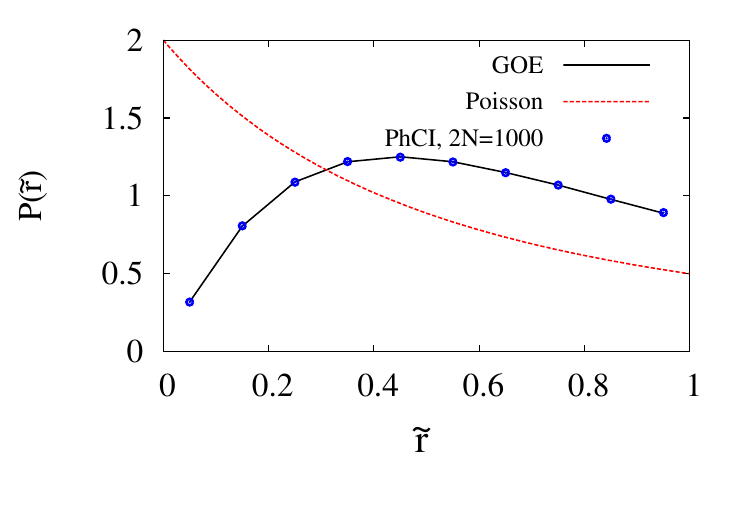}
	\caption{Distribution of nearest neighbor spacing ratio of two kinds for an ensemble of $5000$ Ph-CI matrices of size $2000\times 2000$ }
	\label{figPr}
\end{figure}
The distributions here agree very well with the result of Gaussian 
Orthogonal Ensemble (GOE).

	\section{Density of states}
	\label{DOS}
\begin{figure}[ht!]\centering
		\includegraphics[width=0.38\textwidth]{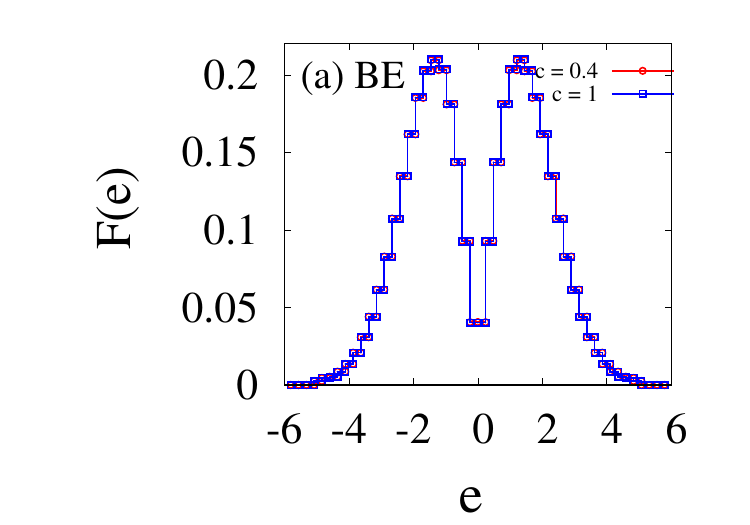}
		\hspace{-1.2cm}
		\includegraphics[width=0.38\textwidth]{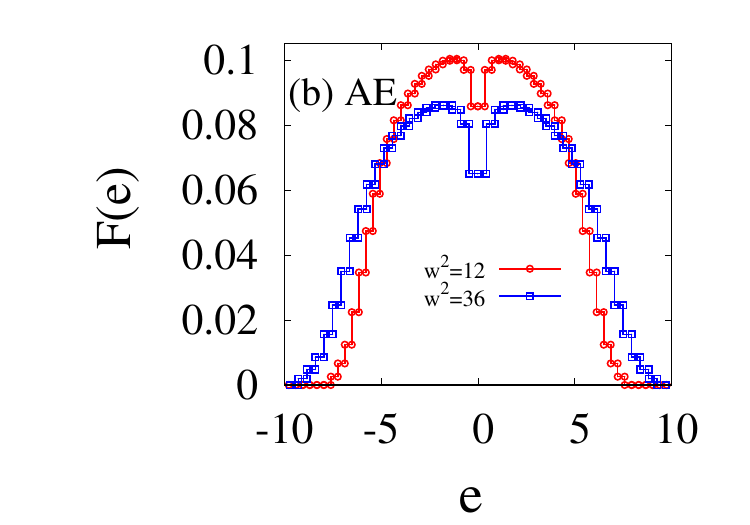}
		\hspace{-1.2cm}
		\includegraphics[width=0.38\textwidth]{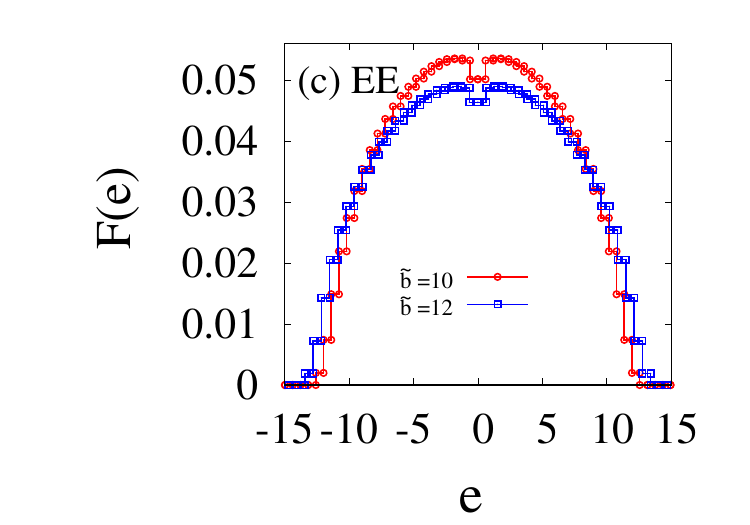}
		\caption{Variance dependence of rescaled density of states 
			$F(e)=\frac{1}{2N}R_1(e)$ for (a) BE, (b) AE and (c) EE. The difference in level density with respect to the ensembles is very much significant. }
		\label{fig1}
	\end{figure}

	The density of states at an energy $e$ is given by 
	$\rho(e)=\sum_n\delta(e-e_n)$ \cite{brody1981random}. 
	The scaled level density $F(e)$ is defined as 
	$F(e)=\frac{1}{2N}R_1(e)$ where $R_1(e)=\langle\rho(e)\rangle$ due to 
	ergodicity in the spectrum \cite{bohigas1975level, brody1981random}. 
	Figure~\ref{fig1} shows that $R_1(e)$ is dependent on the variances of the 
	matrix elements distribution for both AE (\ref{fig1}(b)) and EE 
	(\ref{fig1}(c)) but the dependence is insignificant for BE 
	(\ref{fig1}(a)). This highlights the same finding about sensitivity to 
	ensemble constraints despite being invariant under the same matrix 
	constraint. The exception of BE is again commensurate with earlier findings
	\cite{mondal2020spectral}. The structure of level density curve is highly dependent on the systems obtained by varying ensemble constraints only.

	\section{Energy dependence of complexity parameter}\label{sA}
\begin{figure}[ht!]
		\centering
		\includegraphics[width=0.38\linewidth]{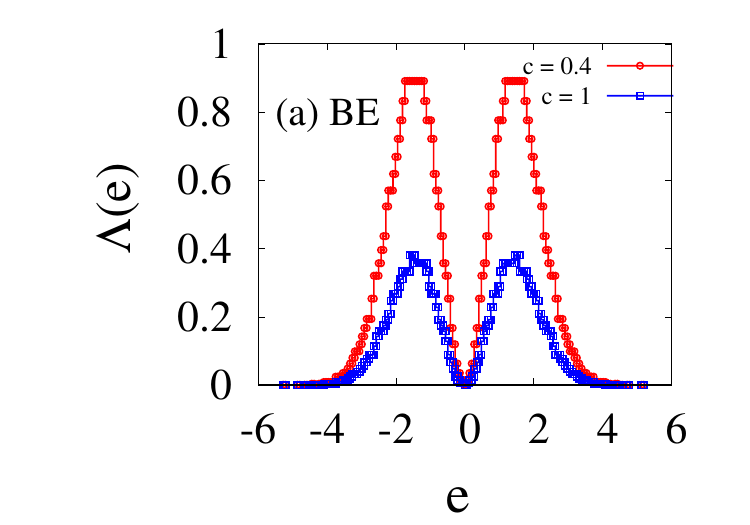}
		\hspace{-1.2cm}
		\includegraphics[width=0.38\linewidth]{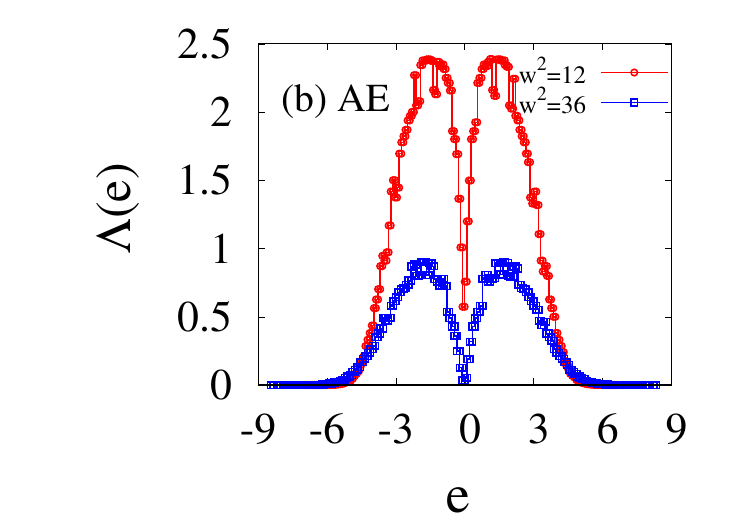}
		\hspace{-1.2cm}
		\includegraphics[width=0.38\linewidth]{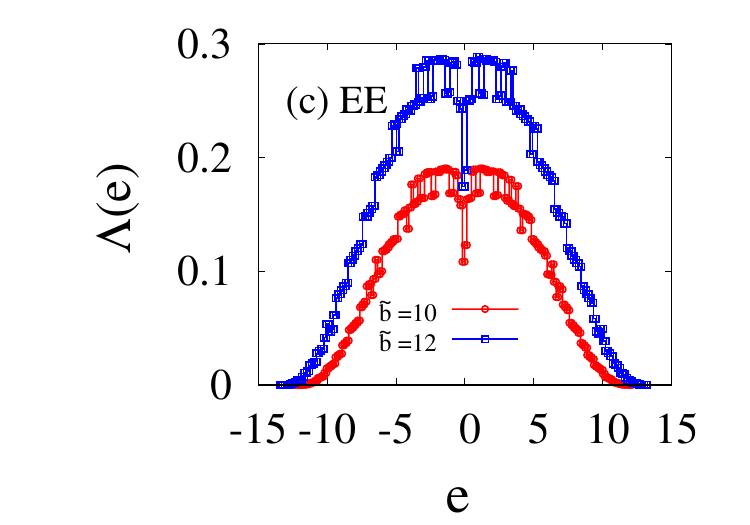}
		\caption{Variation of $\Lambda$ throughout the spectrum for (a) BE, 
			(b) AE and (c) EE for two different variances considered in each case. 
			The spectrum of $\Lambda$ with respect to $e$ is very much dependent 
			on the variances of individual ensembles and also significantly 
			different for different ensembles.}
		\label{fig4}
	\end{figure}

	The dependence of the rescaled complexity parameter $ \Lambda $ on the ensemble constraint variance as a function of energy for three different ensembles is shown in figure~\ref{fig4}. The energy dependence of $ \Lambda $ is provided in its expression in Eq.~\ref{9}.
	
\end{appendices}	

	
	\bibliography{draft}
\end{document}